\documentclass[review,authoryear,12pt]{elsarticle}

\usepackage{graphicx}
\graphicspath{ {./Figures/} }
\usepackage[tight,footnotesize, nooneline]{subfigure}

\usepackage{adjustbox} 
\usepackage[hidelinks]{hyperref} 

\usepackage{amssymb}
\usepackage{amsthm}
\usepackage{amsmath}

\usepackage{lineno}
 
\usepackage{multirow}

\usepackage[flushleft]{threeparttable} 

\journal{arXiv}

\newcommand\copyrighttext{%
   \textcopyright\ 2018 Wiley.}

\begin{document}

\begin{frontmatter}



\title{Transurethral ultrasound therapy of the prostate in the presence of calcifications: a simulation study}





\author[Affil1]{Visa Suomi \corref{cor1}}
\author[Affil2]{Bradley Treeby}
\author[Affil3]{Jiri Jaros}
\author[Affil1]{Pietari Makela}
\author[Affil4]{Mikael Anttinen}
\author[Affil5]{Jani Saunavaara}
\author[Affil5]{Teija Sainio}
\author[Affil1]{Aida Kiviniemi}
\author[Affil1]{Roberto Blanco}

\address[Affil1]{Department of Radiology, Turku University Hospital, Kiinamyllynkatu 4-8, 20521 Turku, Finland}
\address[Affil2]{Department of Medical Physics and Biomedical Engineering, University College London, Gower Street, London, WC1E 6BT, UK}
\address[Affil3]{Centre of Excellence IT4Innovation, Faculty of Information Technology, Brno University of Technology, Bozetechova 1/2, 612 66 Brno, Czech Republic}
\address[Affil4]{Department of Urology, Turku University Hospital, Kiinamyllynkatu 4-8, 20521 Turku, Finland}
\address[Affil5]{Department of Medical Physics, Turku University Hospital, Kiinamyllynkatu 4-8, 20521 Turku, Finland}
\cortext[cor1]{Corresponding Author: Department of Radiology, Turku University Hospital, Kiinamyllynkatu 4-8, 20521 Turku, Finland; Email, visa.suomi@tyks.fi}

\begin{abstract}
\textbf{Purpose:} Transurethral ultrasound therapy is an investigational treatment modality which could potentially be used for the localised treatment of prostate cancer. One of the limiting factors of this therapy is prostatic calcifications. These attenuate and reflect ultrasound and thus reduce the efficacy of the heating. The aim of this study is to investigate how prostatic calcifications affect therapeutic efficacy, and to identify the best sonication strategy when calcifications are present. \textbf{Methods:} Realistic computational models were used on clinical patient data in order to simulate different therapeutic situations with naturally occurring calcifications as well as artificial calcifications of different sizes (1-10~mm) and distances (5-15~mm). Furthermore, different sonication strategies were tested in order to deliver therapy to the untreated tissue regions behind the calcifications. \textbf{Results:} The presence of calcifications in front of the ultrasound field was found to increase the peak pressure by 100\% on average while the maximum temperature only rose by 9\% during a 20-second sonication. Losses in ultrasound energy were due to the relative large acoustic impedance mismatch between the prostate tissue and the calcifications (1.63 vs. 3.20~MRayl) and high attenuation coefficient (0.78 vs. 2.64~dB/MHz$^{1.1}$/cm), which together left untreated tissue regions behind the calcifications. In addition, elevated temperatures were seen in the region between the transducer and the calcifications. Lower sonication frequencies (1-4~MHz) were not able to penetrate through the calcifications effectively, but longer sonication durations (20-60~s) with selective transducer elements were effective in treating the tissue regions behind the calcifications. \textbf{Conclusions:} Prostatic calcifications limit the reach of therapeutic ultrasound treatment due to reflections and attenuation. The tissue regions behind the calcifications can possibly be treated using longer sonication durations combined with proper transducer element selection. However, caution should taken with calcifications located close to sensitive organs such as the urethra, bladder neck or rectal wall.
\end{abstract}


\end{frontmatter}

\copyrighttext

\pagebreak









\section*{Introduction}

Prostate cancer is the second most common cancer occurring in men with an estimated 1.1 million people diagnosed worldwide in 2012~\citep{stewart2014world}. In the same year, approximately 0.3 million people died due to the disease making prostate cancer the fifth most common cause of cancer death among men. To put these figures into perspective, prostate cancer accounts for approximately 15\% of all cancer incidences and 7\% of all cancer related deaths in men~\citep{stewart2014world}. The incidence of prostate cancer increases drastically with age~\citep{zhou2016prostate}, with the vast majority (99\%) of cases diagnosed in men over 50 years old~\citep{stewart2014world}. Patients with prostate cancer may experience non-specific local symptoms as urinary problems or symptoms resulting from distant metastasised disease as bone pain. Prostate cancer may progress locally to adjacent tissues and/or metastasise to other parts of the body including regional and non-regional lymph nodes, bone and extra-nodal soft tissues~\citep{bubendorf2000metastatic}. Therefore, early diagnosis and effective treatment of the disease are essential for the well-being and survival of the patients. 

Typically, clinically significant localised prostate cancer is treated with curative intended therapies such as radical prostatectomy or radiotherapy. These can improve prognosis and survival with or without a multimodal approach depending on disease characteristics. However, they both involve treatment related toxicity that may worsen the outcome of bowel and genitourinary function, possibly impairing quality of life~\citep{fowler1993patient, talcott1998patient}. Therefore, patients would greatly benefit from a treatment option which offers similar therapeutic result with minimal side effects and low risk of post-operative complications. 

Therapeutic ultrasound for the treatment of localised prostate cancer has been studied for over two decades~\citep{madersbacher1995effect} and the clinical results have been promising~\citep{chaussy2001results, thuroff2003high, uchida2006five}. There are no severe side effects from ultrasound if applied appropriately and the treatment can be repeated because there is no cumulative dose limit~\citep{chaussy2001results, poissonnier2007control}. Furthermore, unlike chemotherapy, ultrasound therapy can be used to treat any tumour cell types that are susceptible to high temperatures, which makes it an effective therapy method for multiple cancer types~\citep{sapareto1984thermal}. Therapeutic ultrasound could thus become the primary treatment of localised prostate cancer for selected patients, which offers a minimally invasive alternative for the traditional methods. The treatment has been cleared for clinical use in Europe, Asia and North America, and is currently in use in several hospitals and clinics.

Therapeutic ultrasound treatment of the prostate can be conducted using either a transrectal or transurethral route~\citep{uchida2006five, chin2016magnetic}. This study focuses on transurethral ultrasound therapy, but the results can also be generalised to the transrectal treatment of prostate cancer. The treatment is conducted by first inserting the therapeutic ultrasound probe through the urethra and positioning it near the target. An endorectal cooling device is also used during the therapy in order to prevent the heating and destruction of neighbouring healthy tissues. Once ready, the ultrasound transducer elements in the probe are activated to deliver acoustic energy to the prostate which is absorbed by the tissue creating heat~\citep{chopra2009analysis, siddiqui2010mri}. The probe is automatically rotated in order to ablate the entire tumour volume and magnetic resonance imaging (MRI) is used to monitor and control the heating~\citep{diederich2004transurethral, sammet2015cavernosal}.

The initial clinical evidence from transurethral prostate treatments has shown variability in efficacy~\citep{chin2016magnetic, ramsay2017evaluation}, which could be due to several factors. One factor might be naturally occurring prostatic calcifications~\citep{suh2008calcifications} which can obstruct the penetration of the therapeutic ultrasound field and thus reduce the efficacy of the treatment. The prevalence of prostatic calcifications varies widely from 7\% to 70\% depending on the patient group, and their occurrence and size increases with age~\citep{hyun2018clinical}. Prostatic calcifications are usually classified as being of primary/endogenous type or of secondary/extrinsic type. Endogenous calcifications are thought to be caused by the obstruction of the prostatic ducts around the enlarged prostate by benign prostatic hyperplasia (BPH) or by chronic inflammation. Extrinsic calcifications are caused by urine reflux and thus typically occur around the urethra~\citep{hyun2018clinical}. Prostatic calcifications can occur alone or in aggregates, and their sizes can reach from 0.5~mm to greater than 15~mm~\citep{hong2012prevalence, hyun2018clinical}.

The mechanical and acoustic properties of calcifications differ from those of the prostate tissue, which could potentially cause strong reflections of the ultrasound field. Furthermore, the irregular shape and location of calcifications can deform the ultrasound field making the predictability of the treatment difficult. This effect is further increased by the local aggregation of several calcifications. All these effects can reduce the efficiency of ultrasound therapy in the prostate. For the patient, this  means either longer therapy times, or in the worst case, termination of the treatment. Either case results in inconveniences for both the patient and clinical personnel. The aim of this research is therefore to find out how much the efficacy of ultrasound therapy in the prostate is affected by the scenarios described above. The study is conducted by using both naturally occurring calcifications in segmented computed tomography (CT) patient data as well as artificial calcifications with pre-defined dimensions and locations. The results can be applied to the treatment planning stage in order to improve therapeutic outcomes and to define patient selection criteria for the treatments.

\section*{Treatment simulations}

\subsection*{Study protocol}

Whilst it is difficult, if not impossible, to quantify the effects of calcifications \textit{in vivo} in the prostate, it is possible to replicate their effects using three-dimensional clinical imaging data from patients together with suitable computational models. Hence, the research was carried out using both acoustic and thermal simulation models together with segmented clinical CT data (Optima CT660, GE Healthcare, Chicago, IL, USA / Somatom go.Up, Siemens Healthineers, Erlangen, Germany) from three patients. The patient data were acquired during ultrasound therapy treatments at Turku University Hospital, Finland, using a Tulsa-Pro ultrasound system (Profound Medical Corp, Toronto, Canada). The ultrasound probe of the system has 10 unfocused elements (element size 5~mm $\times$ 4.5~mm with zero spacing) giving a total transducer surface area of 50~mm $\times$ 4.5~mm. The transducer is located 2~mm inside the ultrasound probe which contains a layer of cooling water in front of it. The patients were selected for the simulation study if prostatic calcifications were identified in their CT images. Ethical permission for the study (ETMK: 152/1801/2016) was obtained from the Ethics Committee of Hospital District of Southwest Finland.

\begin{figure}[b!]
\vspace{-0.5cm}
\centering
\includegraphics[height=6.5cm]{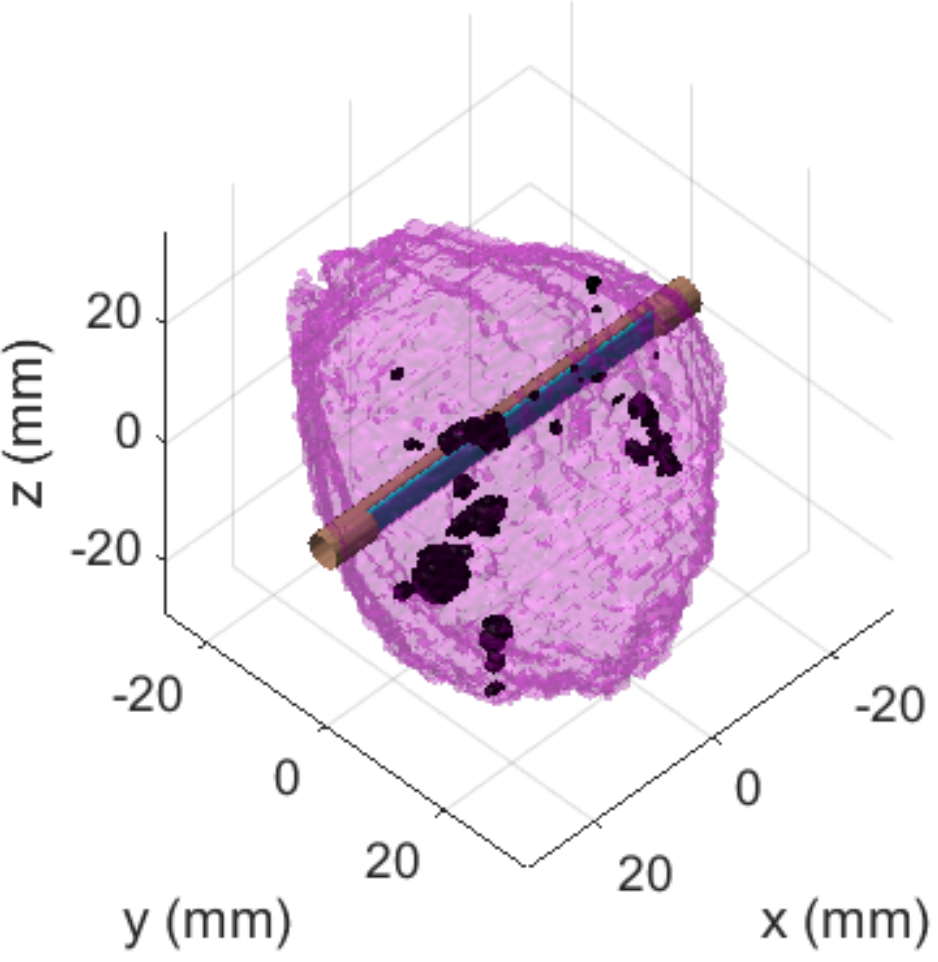}
\caption{The ultrasound probe (brass) was positioned along the urethra in the middle of the prostate (magenta) so that the transducer (cyan) was facing the calcifications (black).}
\label{fig:simulation_geometry}
\end{figure}

The study comprises of three parts. In the first part, data for three different prostate cancer patients are presented. The simulation data comprised of segmented CT volumes of the prostate in each patient. The original resolution of the CT images was 0.97 and 0.95 mm with a slice thicknesses of 1.25 and 1.00 mm for the GE and Siemens CT scanners, respectively. After 3-D interpolation, the grid resolution was set to 0.0462 mm. A visualisation of the simulation geometry is presented in Figure~\ref{fig:simulation_geometry}. Each patient was sonicated using the transurethral ultrasound device (Tulsa-Pro) for 20 seconds followed by a 40-second cooling time. The transducer was operated at 4~MHz and a constant power level was used. The position of the transducer was kept constant throughout the sonication (i.e., no rotation) so that it was facing the calcifications. The results compare the acoustic pressure parameters as well as the evolution of temperature and thermal dose in each individual patient.

In the second part of the study, artificial calcifications of different sizes and locations were placed in the prostate of the same patient in order to examine their effect on therapeutic efficacy. In each case, the artificial calcification had a spherical shape, with its diameter varied between 1-10 mm and its location (the centre point) from the transducer face varied between 5-15~mm. Different sonication durations (10-40~s) and frequencies (1-4~MHz) with a fixed calcification diameter (5~mm) and location (10~mm) were also used to study their effect on the measured parameters. The same power level was used in all cases.

In the third part of the study, different sonication strategies were tested using clinical patient data in order to treat the tissue region behind the calcification. The number of active elements and the sonication durations of individual elements were changed in order to ablate the target region. The best strategy to treat the tissue region behind the calcification was then suggested based on the results.

\subsection*{Acoustic simulation parameters and execution}

A more detailed description of the acoustic simulation model can be found from a previous study~\citep{suomi2017full}, but it is shortly described here. The ultrasound simulations were carried out using the parallelised C++ version of the open source k-Wave Toolbox~\citep{treeby2012modeling, jaros2016full}. The code solves a set of three coupled first-order partial differential equations. These are based on the conservation of mass, conservation of momentum, and a pressure-density relation that includes a phenomenological loss term accounting for acoustic absorption that follows a frequency power law. The governing equations are equivalent to a generalised version of the Westervelt equation that accounts for second-order acoustic nonlinearity, power law acoustic absorption, and a heterogeneous distribution of material properties (sound speed, density, nonlinearity and absorption coefficient).

The ultrasound simulations were run on a computing cluster at CSC - IT Centre for Science, Finland, using 384 cores (Haswell / Sandy bridge, Intel, Santa Clara, CA, USA), up to 150~GB memory and approximately 4~hours per simulation. Several convergence studies were run prior the study in order to define the optimal grid size and temporal resolution for accuracy and computational speed. The sizes of the computational grids were at maximum 1152 $\times$ 1536 $\times$ 384 grid points, i.e., 5.3~cm $\times$ 7.1~cm $\times$ 1.8~cm, and they supported harmonic frequencies up to 16~MHz (i.e., four harmonics with the sonication frequency of 4~MHz). The temporal resolution was set to 30 points per period (at 4~MHz) which corresponded to a time step of 8.3~ns. The last three cycles of the time domain waveforms were saved over the whole grid for data analysis.

The acoustic properties of tissues used in the simulations are presented in Table \ref{tab:acoustic_parameters}~\citep{mast2000empirical, hasgall2015database, parker1993elastic}. Since the acoustic properties of prostatic calcifications are not characterised in the literature, they were defined based on materials whose chemical composition is similar. Renal stones consist mainly of calcium oxalate~\citep{manglaviti2011vivo} which is also one of the main components of prostatic calcifications~\citep{sfanos2009acute, dessombz2012prostatic}. Therefore, it is reasonable to define the acoustic properties of prostatic calcifications as those of renal stones~\citep{singh1989study}. Due to the large variation in the measured values, the average value for each property was used. If the variation is a result of natural differences in the calcifications rather than measurement inaccuracy, this could possibly create local changes to the acoustic field within the calcifications in the simulations.

\begin{table}[t!]
  \centering
  \caption{Acoustic simulation parameters~\citep{mast2000empirical, hasgall2015database, parker1993elastic}}
    \begin{tabular}{lcccc}
    \hline
    \hline
          			& Density   		& Sound speed   	& Attenuation 				& B/A 	\\
          			& (kg/m$^{3}$) 		& (m/s) 			& (dB/MHz$^{1.1}$/cm)		& 		\\
    \hline
    Prostate		& 1045  			& 1561				& 0.78			   			& 7.5 	\\
    Calcification	& 1466  			& 2184				& 2.64			   			& 7.5 	\\
    Muscle	 		& 1050  			& 1547  			& 1.09   					& 7.2 	\\
    Fat   			& 950   			& 1478  			& 0.48  					& 10.1 	\\
	Water 			& 1000  			& 1520  			& 0.00217 					& 5.2 	\\
    \hline
    \hline
    \end{tabular}
  \label{tab:acoustic_parameters}
\end{table}

\subsection*{Thermal simulation parameters and execution}

The thermal simulations were conducted using the k-Wave Toolbox. This solves Pennes bioheat transfer equation using a non-standard pseudospectral method~\citep{treeby2017nonstandard}. The model is exact in the case of homogeneous media, and gives high accuracy for low computational cost in the case of heterogeneous material parameters. The model accounts for heat diffusion, advective heat loss due to tissue perfusion and heat deposition due to nonlinear ultrasound absorption. The solution took into account the specific heat capacity, thermal conductivity and the perfusion in different tissues. The heating rate was calculated using the harmonic components of the nonlinear ultrasound field. This was done in order to accurately replicate the increased heating effect of the ultrasound field due to nonlinearity.

The thermal simulations were run in Matlab R2017a (MathWorks, Natick, Massachusetts, United States) on same cluster as the acoustic simulations with approximately 4 hours per simulation. The grid resolution was decimated by a factor of 4 (with respect to ultrasound simulations) for computational efficiency and a time step of 20~ms was used. The temperature of the water inside the ultrasound probe was held constant at 21~$^{\circ}$C to mimic the cooling effect of the room temperature water flowing through the clinical system. The perfusion rate was set to zero for tissue regions which reached a thermal dose of 240 cumulative equivalent minutes at 43~$^{\circ}$C (CEM).

The thermal simulations were conducted using the tissue parameters in Table~\ref{tab:thermal_parameters}~\citep{van2002prostate, patch2011specific, hasgall2015database}. The thermal properties of prostatic calcifications are likewise not characterised in the literature, and thus the thermal properties of calcium carbonate were used~\citep{ramachandran2006photoacoustic}. Calcium carbonate crystals promote calcium oxalate crystallisation by heterogeneous or epitaxial nucleation~\citep{geider1996calcium}, which is why it is often found in prostatic calcifications alongside with calcium oxalate~\citep{sfanos2009acute, dessombz2012prostatic}.

\begin{table}[t!]
  \centering
  \caption{Thermal simulation parameters~\citep{van2002prostate, patch2011specific, hasgall2015database}}
    \begin{tabular}{lccc}
    \hline
    \hline
          			& Thermal			& Specific		& Perfusion 		\\
          			& conductivity		& heat capacity	& rate				\\
          			& (W/m/K) 			& (J/kg/K) 		& (kg/m$^{3}$/s)	\\
    \hline
    Prostate		& 0.51			 	& 3400 			& 1.7				\\
    Calcification	& 3.85			 	& 858 			& 0					\\
    Muscle			& 0.49 				& 3421  		& 0.6				\\
    Fat				& 0.21				& 2348			& 0.6				\\
    Water			& 0.60				& 4178			& 0					\\
    Blood 			& N/A      			& 3617  		& N/A 				\\
    \hline
    \hline
    \end{tabular}
  \label{tab:thermal_parameters}
\end{table}

\section*{Results}

\subsection*{Study with natural calcifications}

\begin{figure}[b!]
\centering
\subfigure[]{\includegraphics[height=4cm]{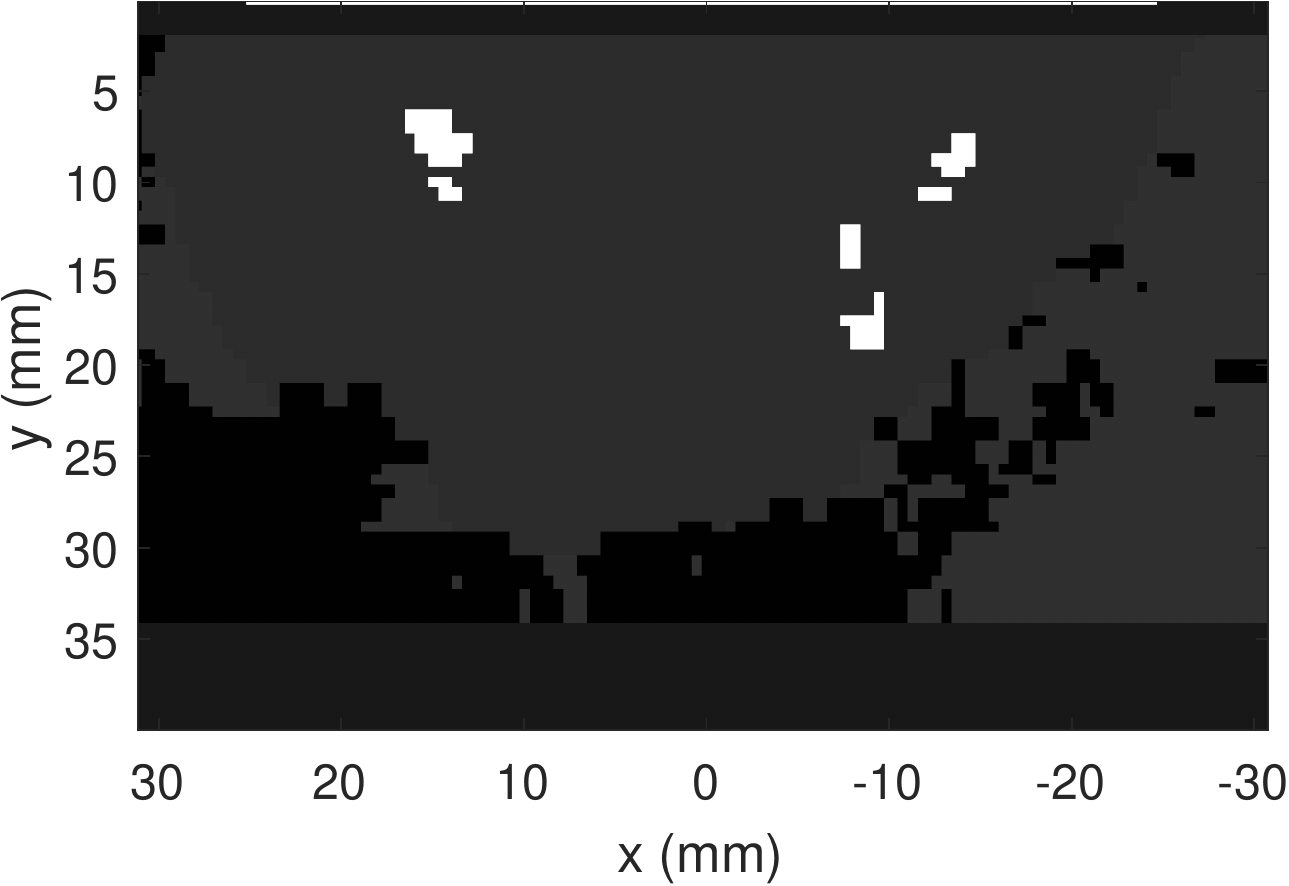}}
\subfigure[]{\includegraphics[height=4cm]{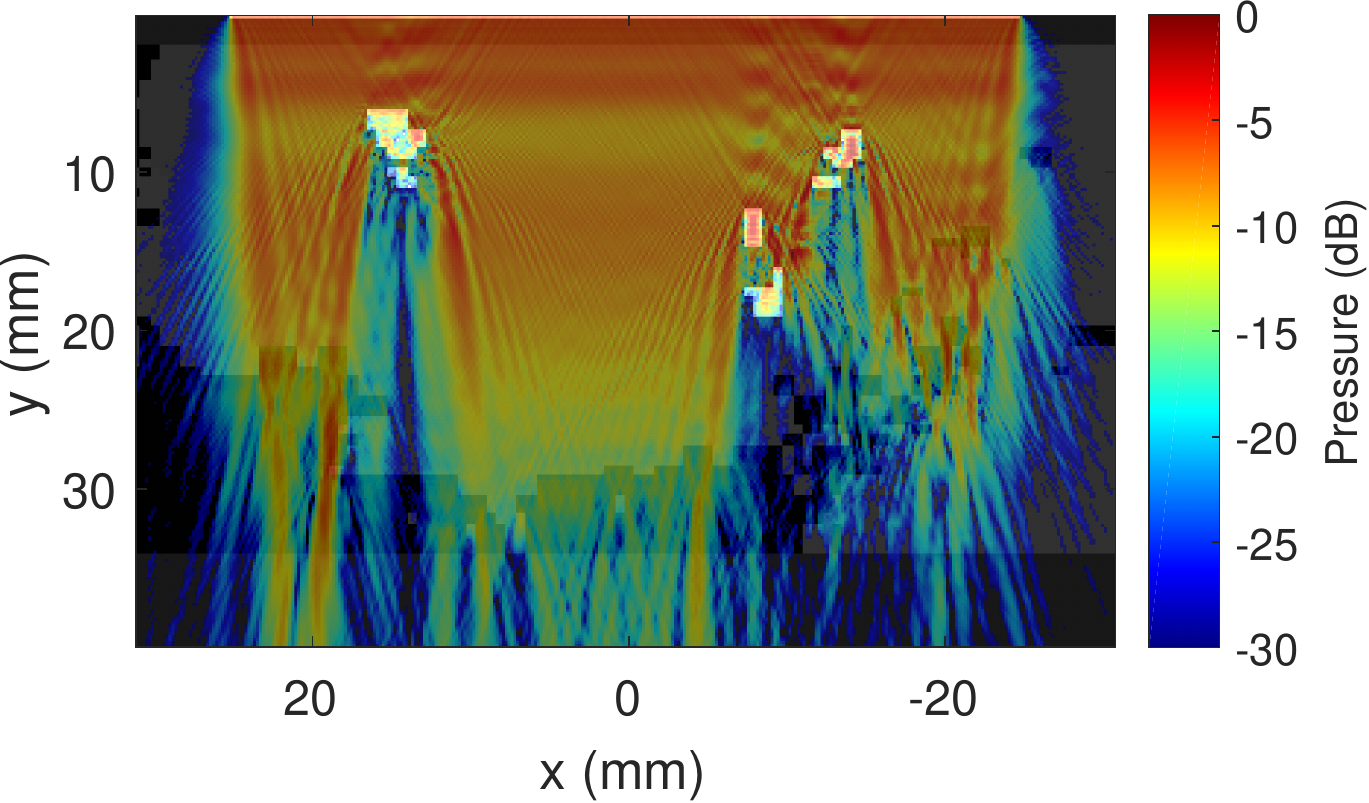}}
\\
\subfigure[]{\includegraphics[height=4cm]{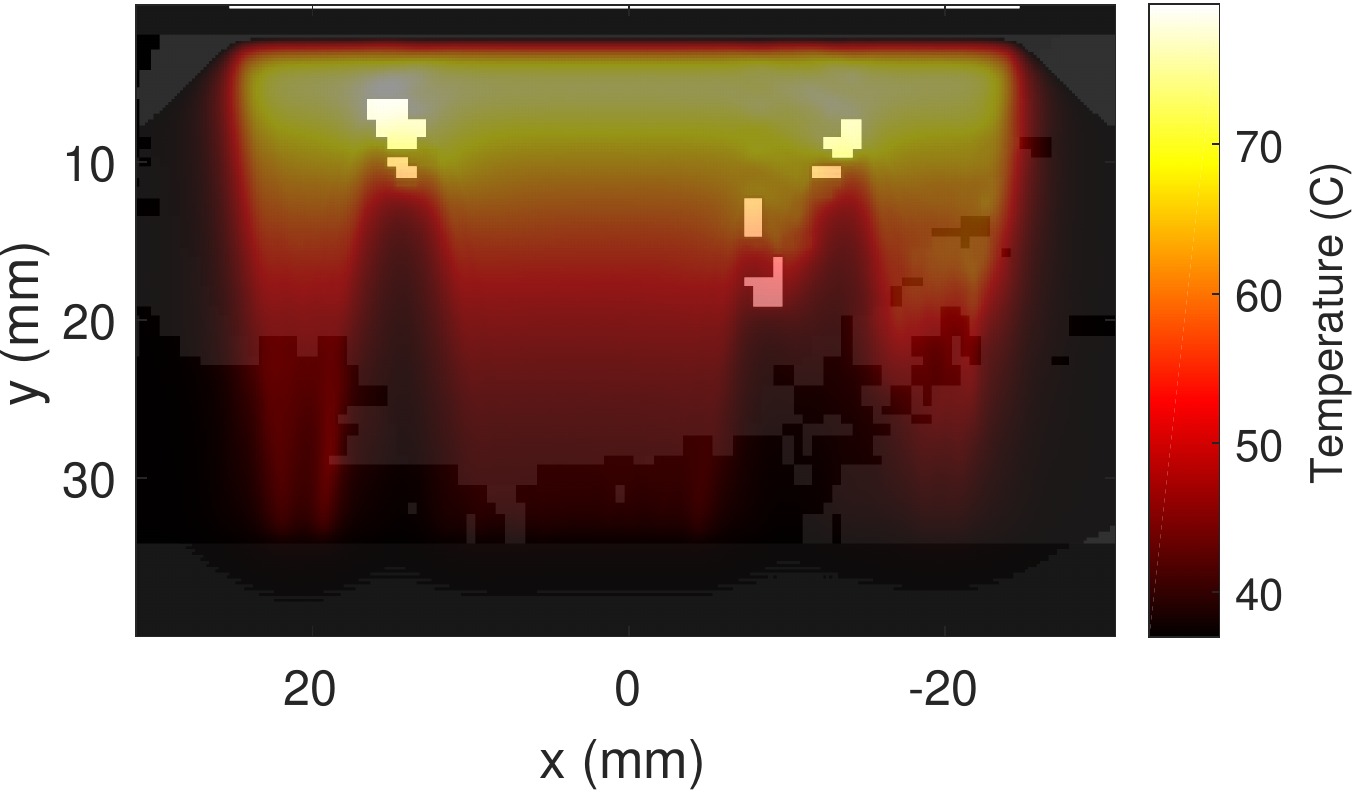}}
\subfigure[]{\includegraphics[height=4cm]{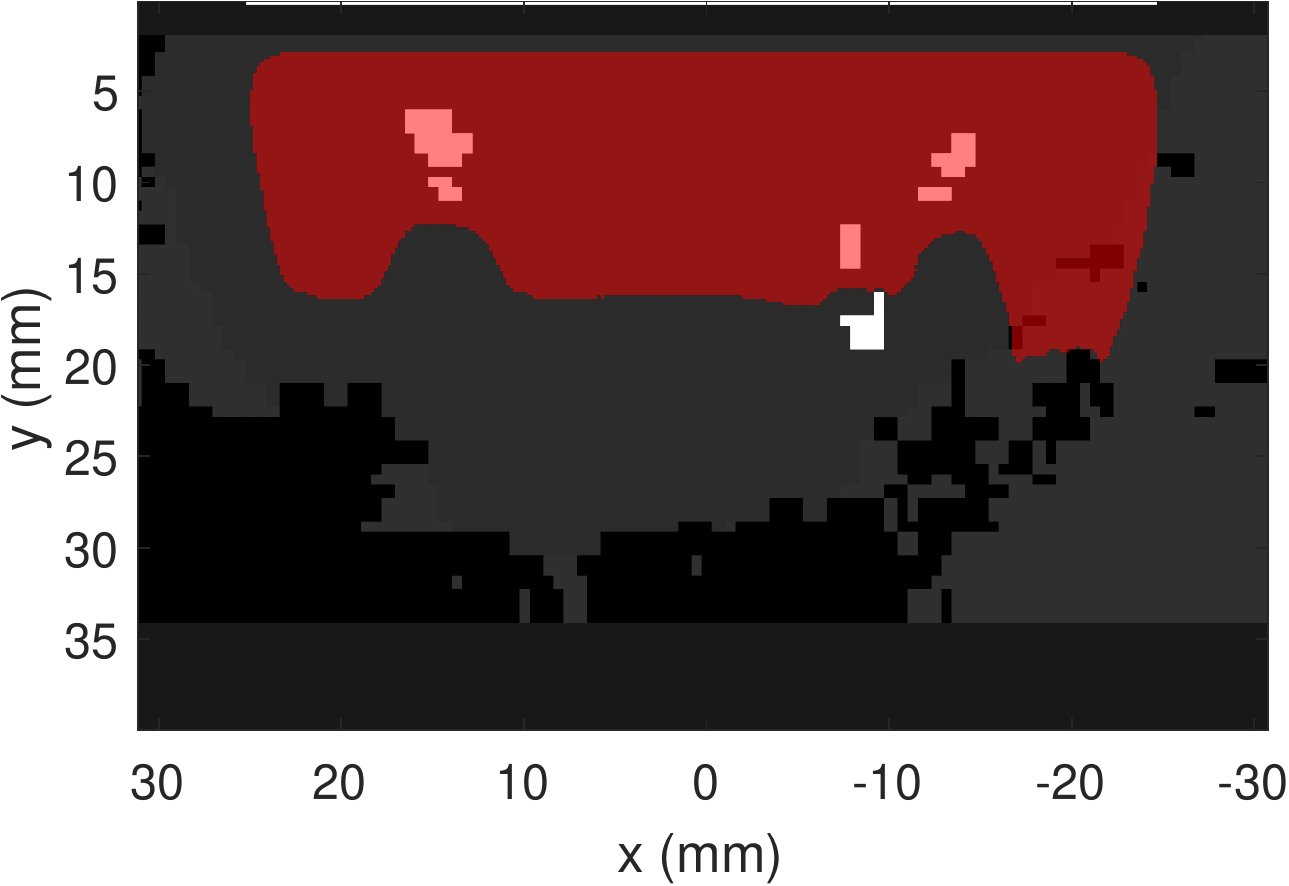}}
\caption{Segmented computed tomography (CT) slice of the half of the prostate (grey) with calcifications (white) in patient 1. The tissue regions  surrounding the prostate were segmented as fat (black) and muscle (light grey). The ultrasound probe was positioned along the urethra in the middle of the prostate (i.e., on the top of the image). Simulated (b) ultrasound, (c) temperature and (d) 240 cumulative equivalent minutes (CEM) thermal dose fields at the end of the sonication.}
\label{fig:natural_planes}
\end{figure}

Figure~\ref{fig:natural_planes}(a) shows an axial slice of the segmented CT data from patient 1 along the centre of the transducer (the x-axis represents the craniocaudal direction). The tissue regions in the image are prostate (grey), muscle (light grey), fat (black) and calcifications (white). Figure~\ref{fig:natural_planes}(b) shows the same CT slice together with the simulated ultrasound field. The peak pressure of the ultrasound field in this case was 2.1 MPa which was located in front of the large calcification at x = 15~mm, y = 7~mm. The calcifications cause strong reflection and attenuation of the ultrasound field. This means very little acoustic energy reaches the regions behind the calcifications. The strong reflections are created by the large acoustic impedance differences between the prostate tissue and the calcifications (1.63 vs. 3.20~MRayl), which causes approximately 10.6\% of the incident energy to be reflected at the interface. In addition, the incident ultrasound energy is further decreased inside the calcifications due to their high attenuation coefficient (0.78 vs. 2.64~dB/MHz$^{1.1}$/cm).

Figure~\ref{fig:natural_planes}(c) shows the temperature field on top of the CT slice at the end of a 20-second sonication. The peak temperature at the end of the sonication was 79.8~$^{\circ}$C which was located at x = 15~mm, y = 5~mm. Again, it can be seen that the temperature does not increase drastically behind the calcifications apart from the regions located in close proximity to the calcifications. On the contrary, the temperature inside the calcifications rises faster than the rest of the prostate tissue due to their relatively low specific heat capacity (3400 vs. 858~J/kg/K) and lack of perfusion. However, this thermal energy is also dissipated away quickly due to the high thermal conductivity (0.51 vs. 3.85~W/m/K). This causes the tissue regions surrounding the calcifications to absorb some of the thermal energy. Figure~\ref{fig:natural_planes}(d) shows the 240 CEM thermal dose field after the 20-second sonication followed by a 40-second cooling time. The thermal damage to the tissue regions behind the calcifications does not extend as far as in the regions where there are no calcifications in the way of the ultrasound field. Only the tissues which are in direct contact with calcifications have absorbed enough thermal energy to induce damage. Some thermal damage has also been induced to the tissues outside the prostate on the right-hand side of the figure.

\begin{figure}[b!]
\vspace{-0.5cm}
\centering
\subfigure[]{\includegraphics[height=4cm]{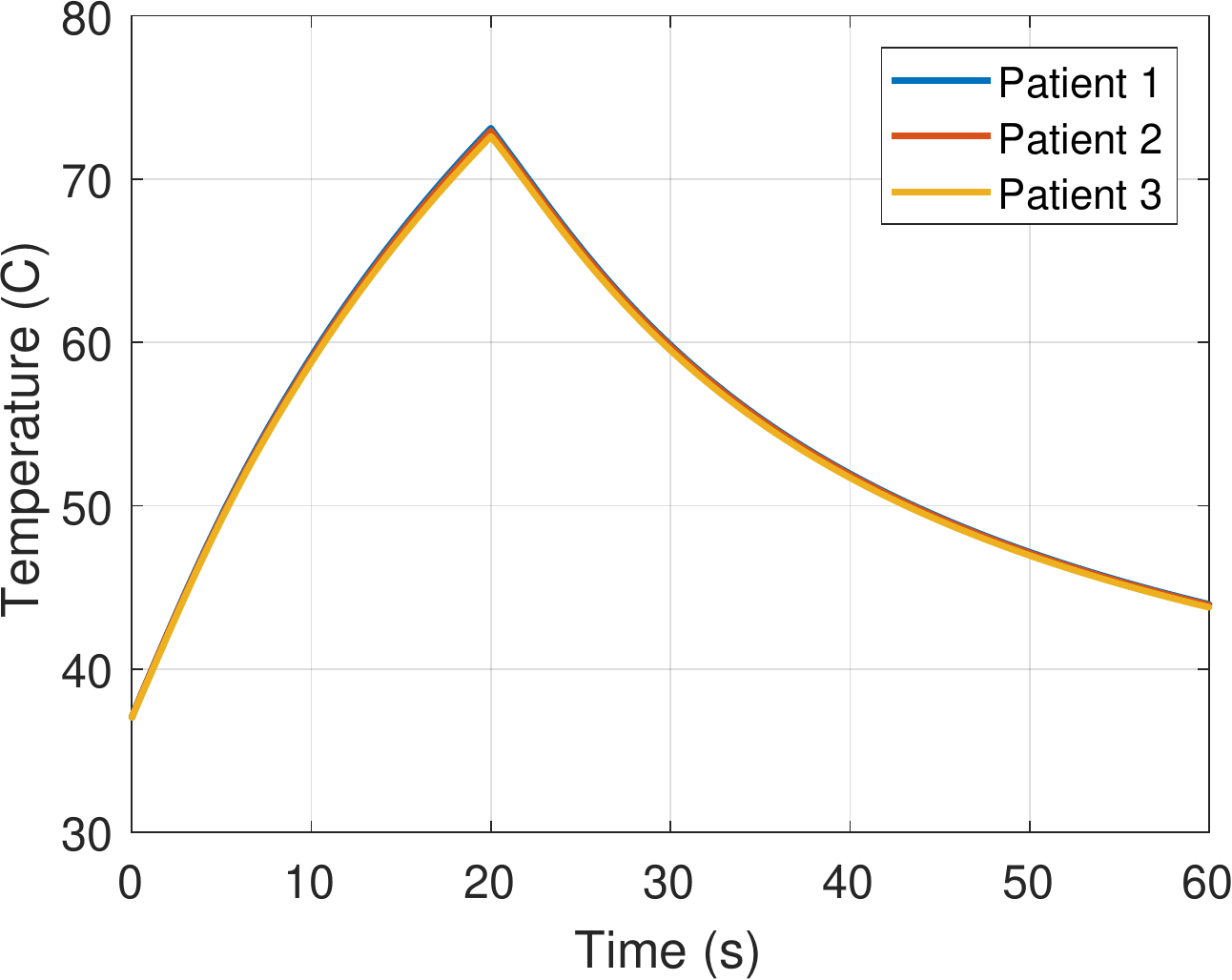}}
\subfigure[]{\includegraphics[height=4cm]{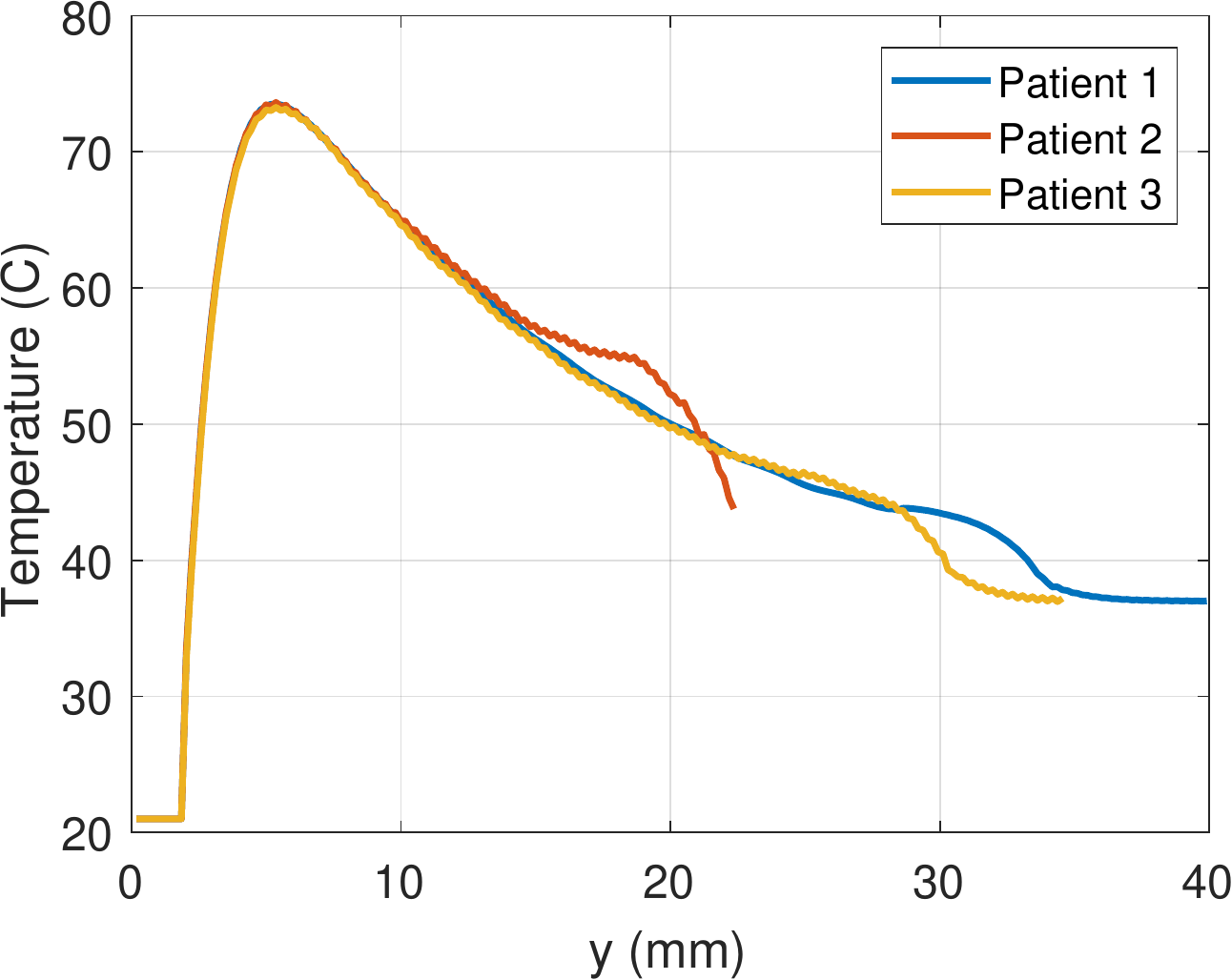}}
\subfigure[]{\includegraphics[height=4cm]{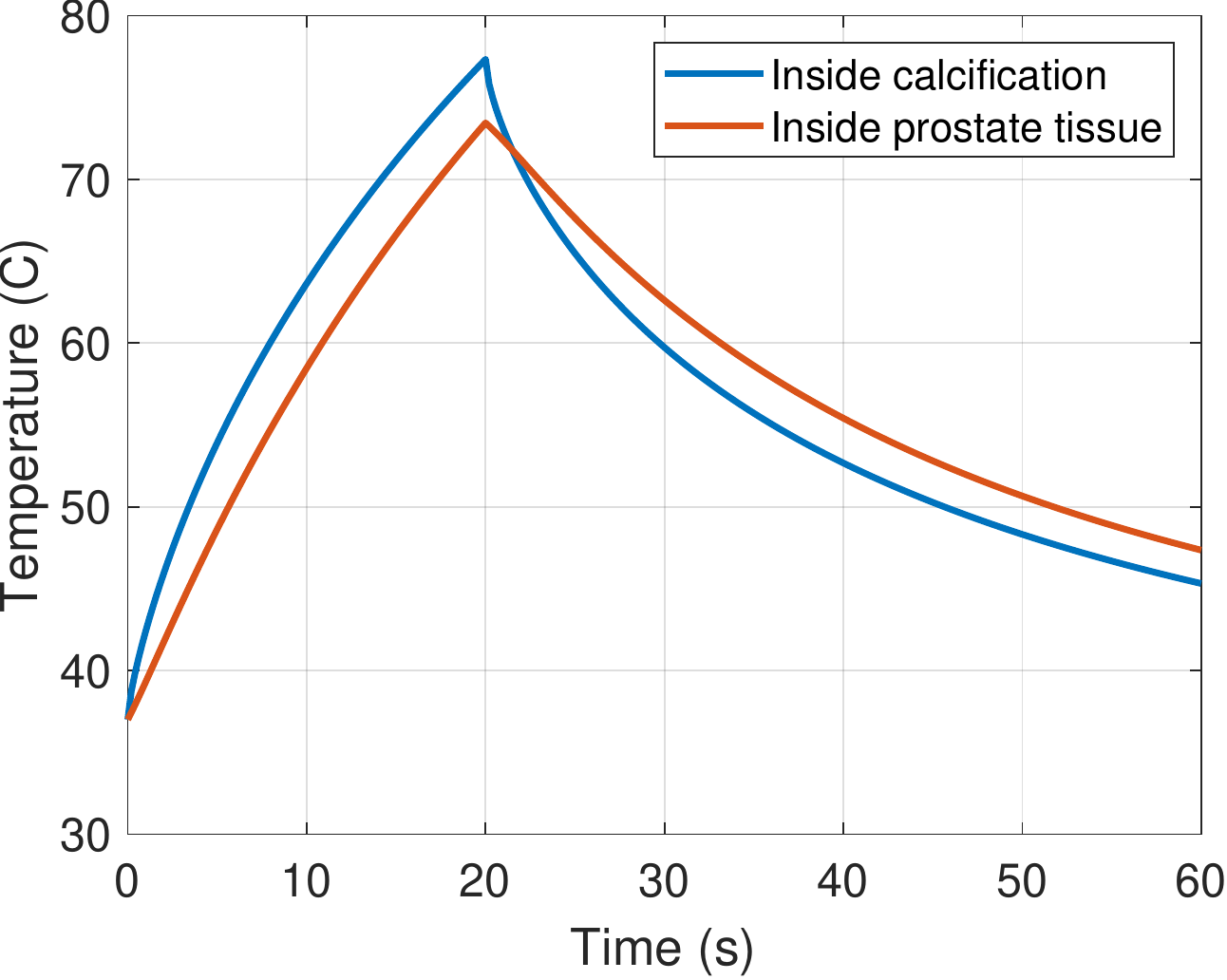}}
\caption{(a) Temperature evolution measured at a distance of 5~mm from the centre of the transducer during a 20-second sonication followed by a 40-second cooling time in three patients. (b) Axial temperature profiles at the end of the 20-second sonication along the y-axis from the centre of the transducer (the origin). The ultrasound probe was held at a constant temperature of 21~$^{\circ}$C. (c) Evolution of the temperature inside the calcification with respect to the prostate tissue at the same distance from the transducer in patient 2.}
\label{fig:natural_temp_profiles}
\end{figure}

The temperature evolution at 5~mm from the centre of the transducer (the origin) in all patients is presented in Figure~\ref{fig:natural_temp_profiles}(a). The temperature profiles at these locations are in the prostate tissue (i.e., not inside the calcifications). It is evident from the figure there are no drastic differences between the three patients, with the temperatures reaching a maximum of approximately 73~$^{\circ}$C in each case after the 20-second sonication. In the clinical setting, a minimum temperature of 55~$^{\circ}$C is desired inside the treatment region. The axial temperature profiles (along the y-axis) from the centre of the transducer are shown in Figure~\ref{fig:natural_temp_profiles}(b). The temperature inside the transducer was kept constant at 21~$^{\circ}$C, which is shown as a horizontal line along the first 2~mm from the transducer. Again, big differences are not seen between different patients, with temperatures peaking approximately at y = 5.4~mm. After the peak, the temperature curves decrease steadily towards the outer edge of the prostate where steeper drops are seen.

Figure~\ref{fig:natural_temp_profiles}(c) shows the temperature evolution inside the calcification with respect to the prostate tissue at the same distance in patient 2. It can be seen that the temperature rises faster inside the calcification due to the lower specific heat capacity but also dissipates faster after the sonication due to the higher thermal conductivity. This happens despite the fact that there is no perfusion in the calcification.

\begin{figure}[htbp!]
\centering
\subfigure[]{\includegraphics[height=4cm]{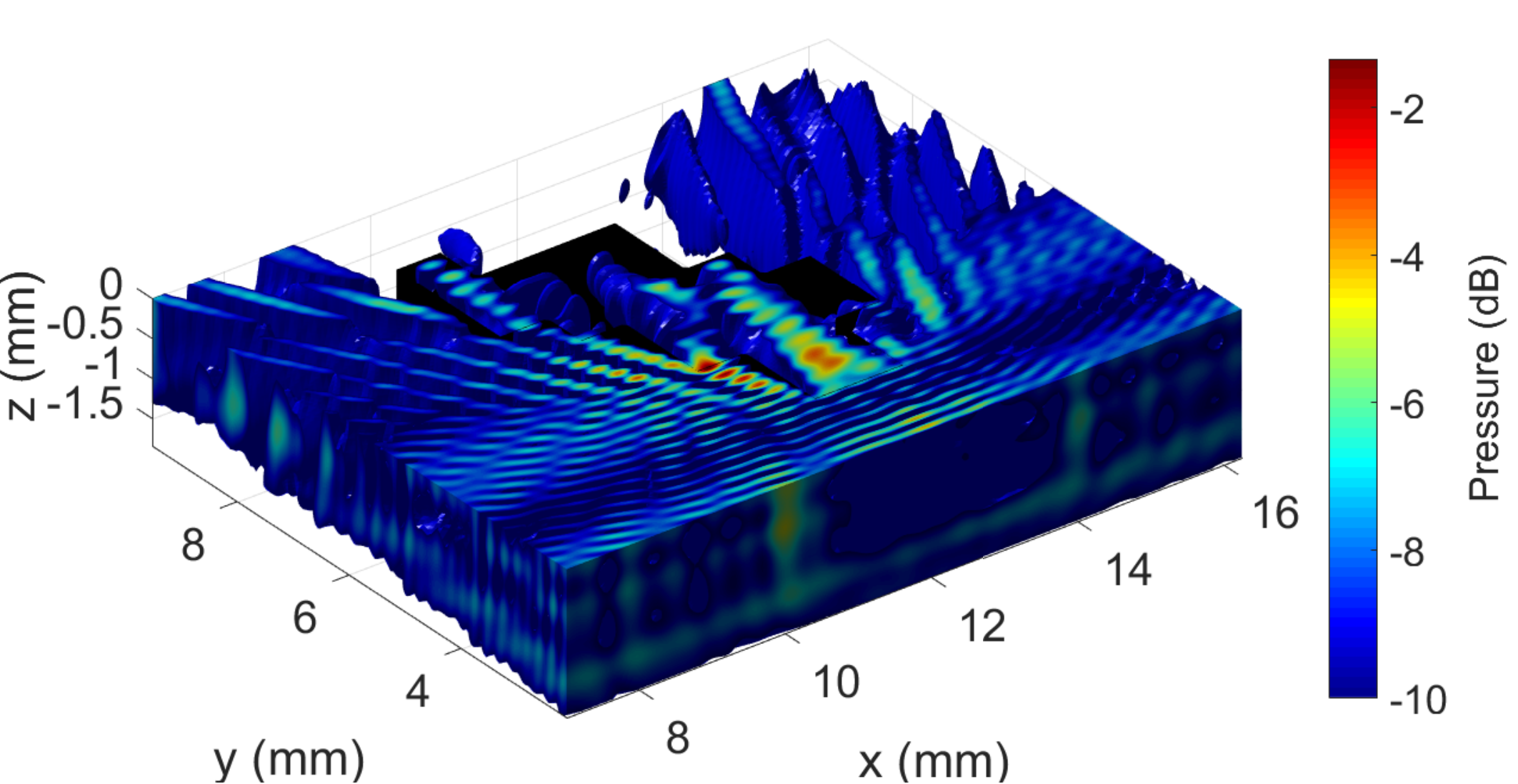}}
\subfigure[]{\includegraphics[height=4cm]{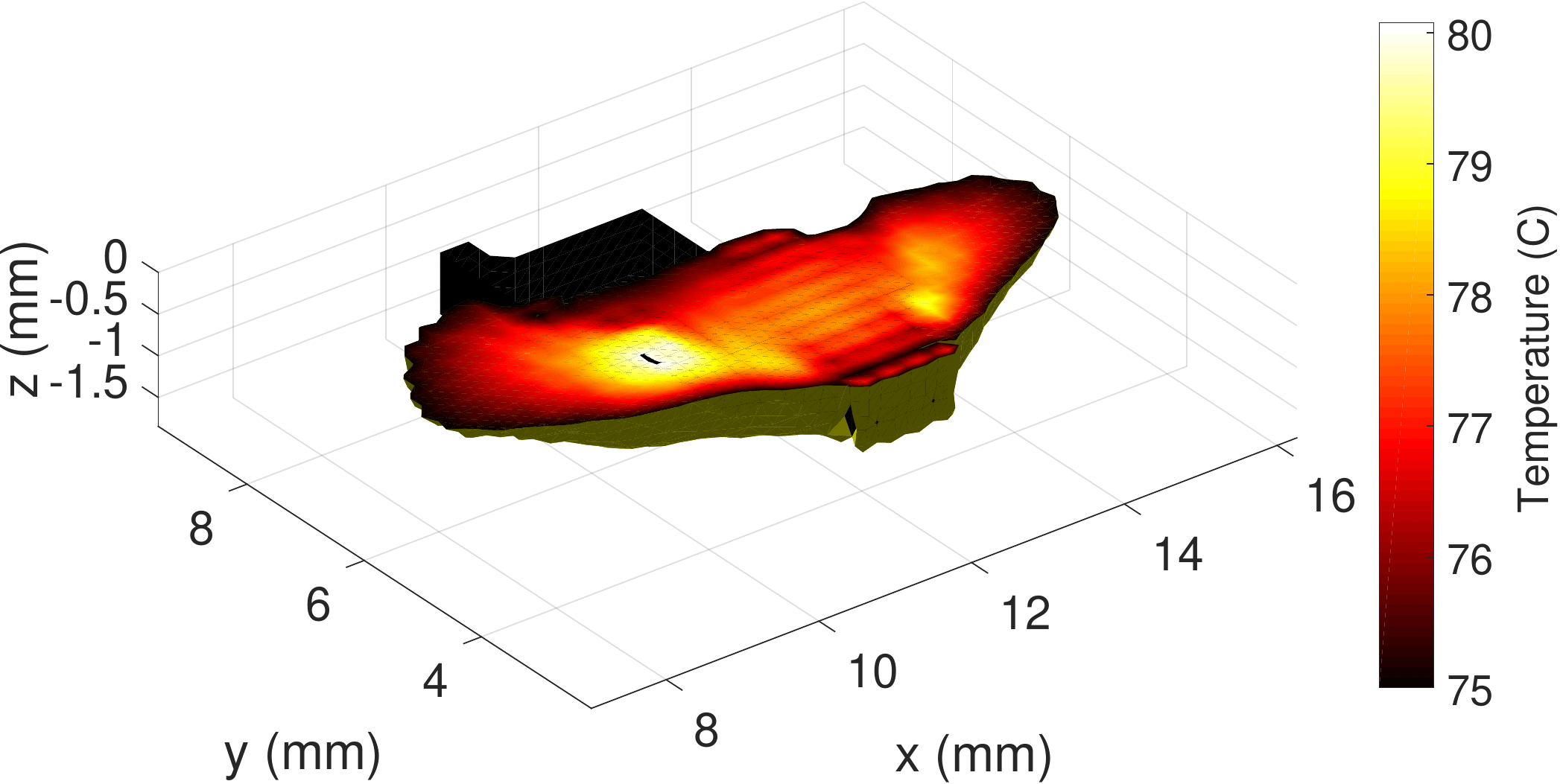}}
\subfigure[]{\includegraphics[height=4cm]{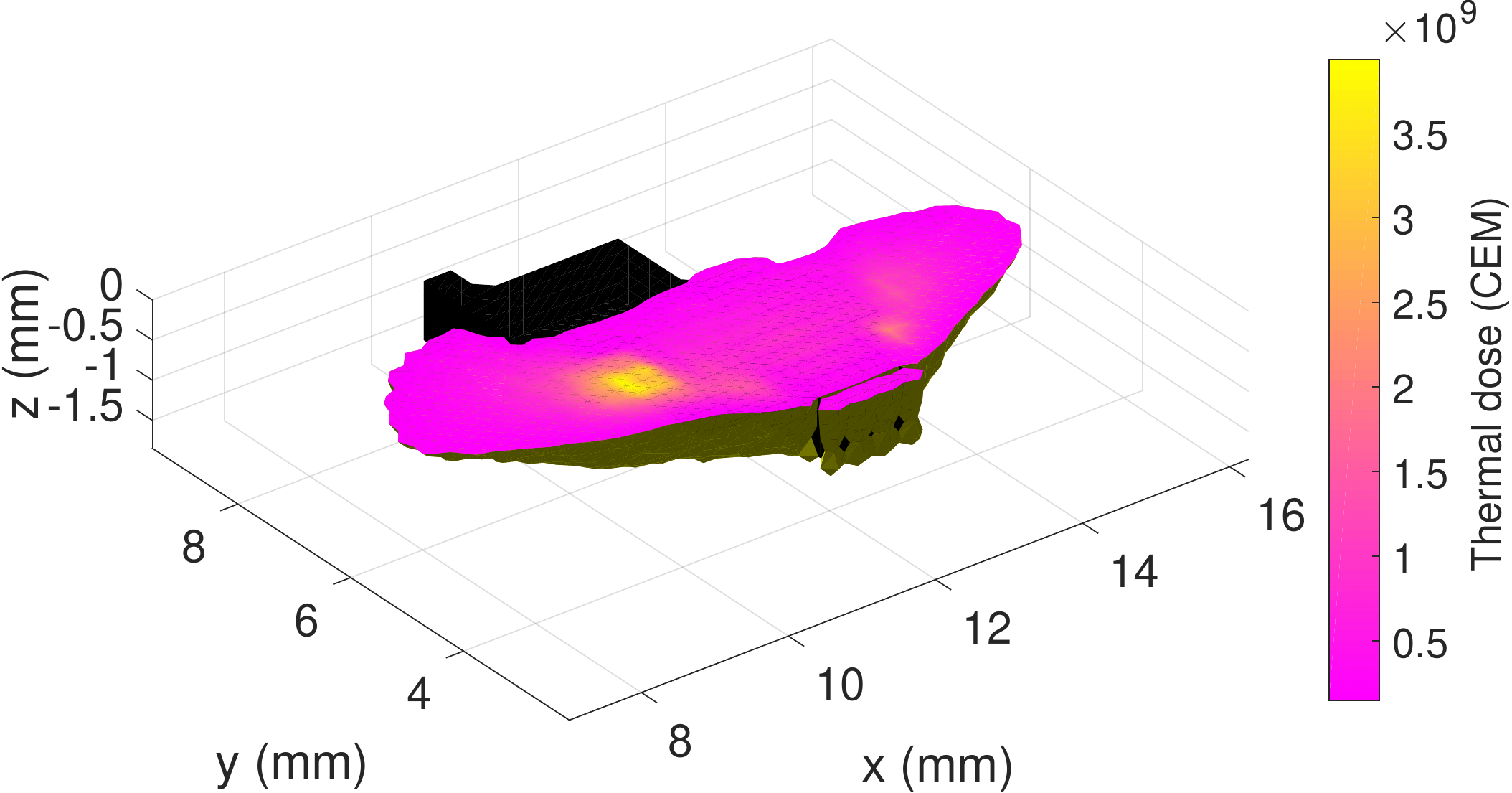}}
\caption{Three-dimensional visualisation of (a) the acoustic pressure field (thresholded at $-$10~dB) and the evolution of (b) temperature (thresholded at 75~$^{\circ}$C) and (c) thermal dose (thresholded at 1.5$\times$10$^{8}$~CEM) fields in the presence of a calcification (black) inside the prostate during a 20-second sonication in patient 2.}
\label{fig:natural_isocaps}
\end{figure}

A close up cutout of the distortion of the ultrasound field (thresholded at $-$10~dB) in the presence of a calcification is shown in Figure~\ref{fig:natural_isocaps}(a) in patient 2. A strong pressure region is formed in front of the calcification which is also partly penetrates inside. As mentioned earlier, the strong reflections are caused by the large acoustic impedance differences between the prostate and the calcifications. The ultrasound energy is further attenuated inside the calcification due to its high attenuation coefficient and then reflected again when entering the prostate from the back side. The temperature and thermal dose field cutouts during the 20-second sonication are presented in Figures~\ref{fig:natural_isocaps}(b)-(c), respectively. The temperature and thermal dose fields start to evolve first in the vicinity of the calcification where the strongest pressure regions are located. The heating rate is directly related to the intensity of the ultrasound field which shows as high temperature and thermal dose values in the same locations. The high thermal conductivity of the calcification also dissipates heat energy to the tissue regions surrounding the calcification.

\begin{figure}[b!]
\vspace{-0.5cm}
\centering
\subfigure[]{\includegraphics[height=4cm]{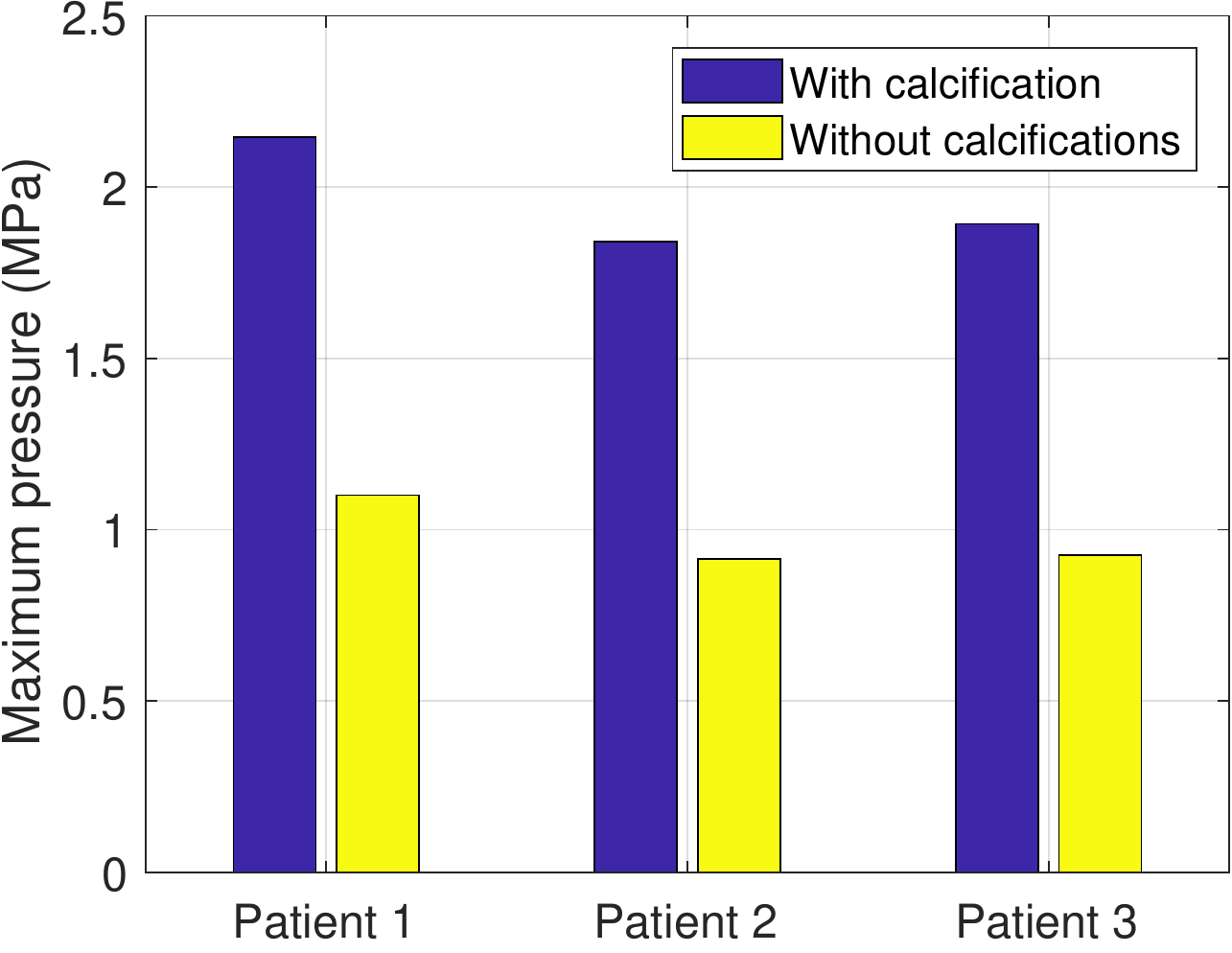}}
\subfigure[]{\includegraphics[height=4cm]{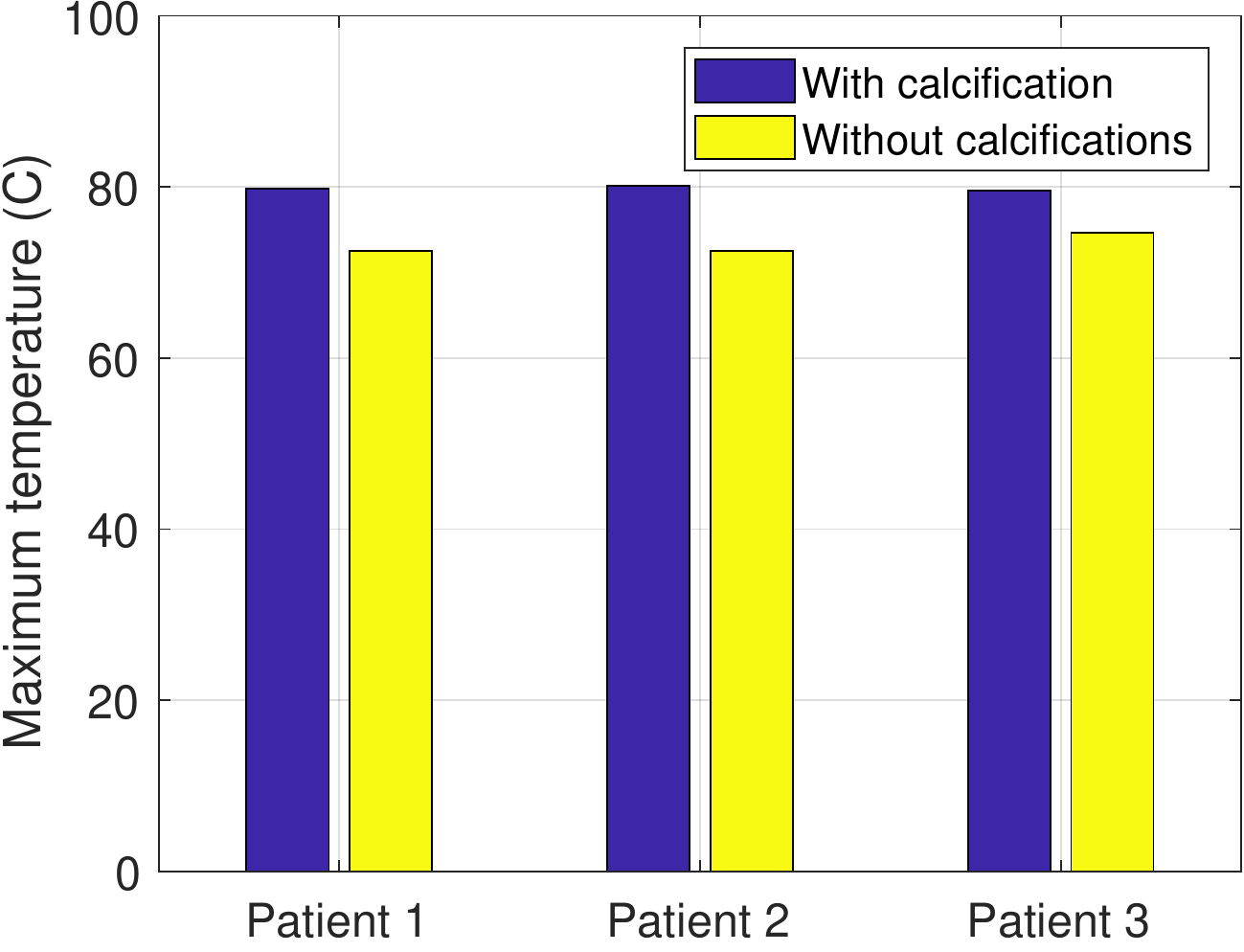}}
\subfigure[]{\includegraphics[height=4cm]{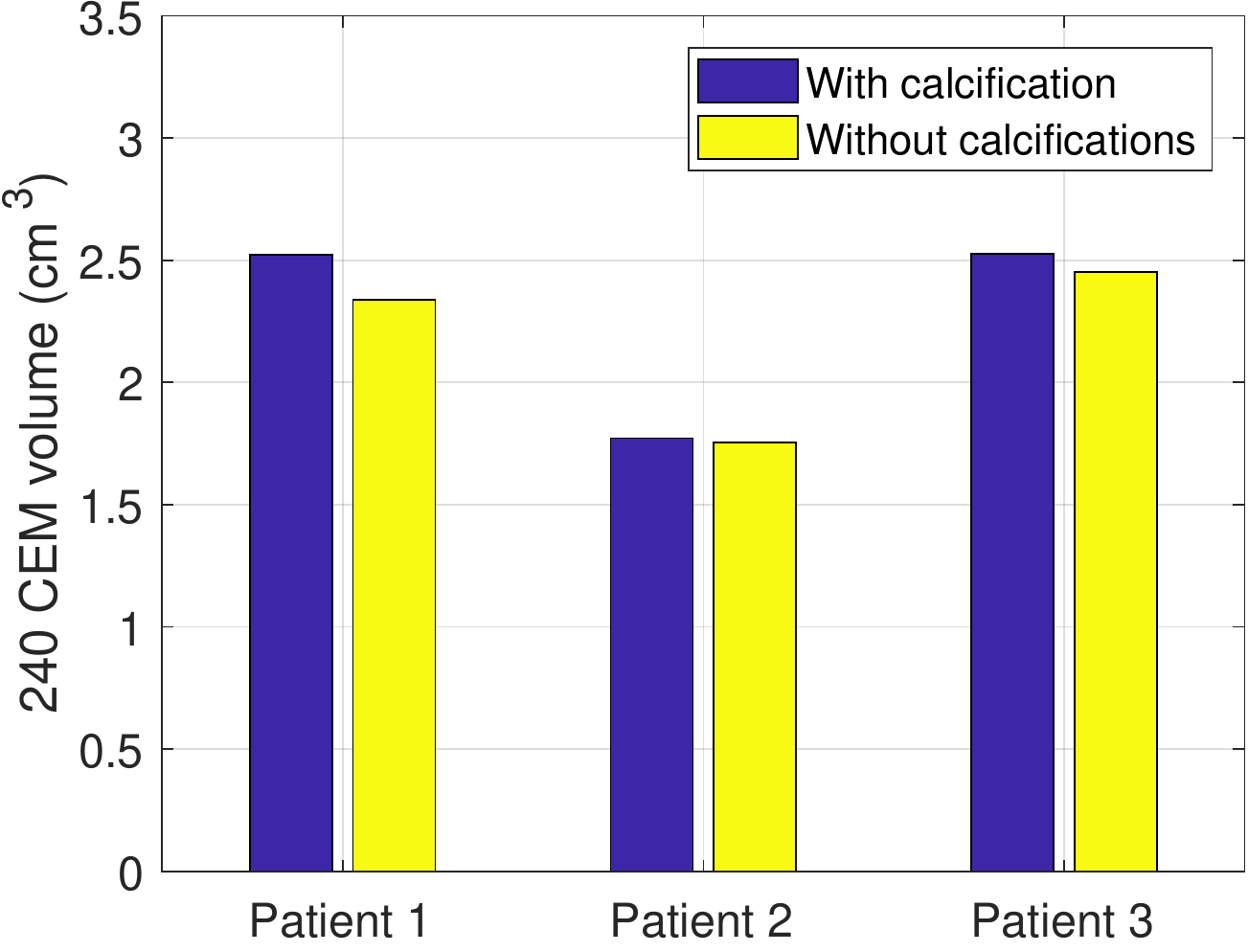}}
\caption{Comparison of (a) maximum acoustic pressures, (b) maximum temperatures and (c) 240~CEM thermal dose volumes in three different patients with and without naturally occurring calcifications. The sonication duration was 20 seconds followed by a 40-second cooling time.}
\label{fig:natural_bar}
\end{figure}

Comparison of the maximum acoustic pressures, temperatures and thermal dose volumes in all patients are shown in Figures~\ref{fig:natural_bar}(a)-(c), respectively. The maximum pressures in the presence of calcifications are two-fold (99.9\% higher on average) compared to the simulations without calcifications. As mentioned earlier, the high pressures are caused by the large acoustic impedance differences between the calcification and the prostate tissue, which causes the ultrasound field to strongly reflect at the interface causing the accumulation of the acoustic pressure. The temperatures in the presence of calcifications are only slightly (9.1\% on average) higher when compared to the same prostate without calcifications. Although the regions with calcifications tend to heat up faster due to the presence of high pressure regions as well as their relatively low specific heat capacity, they also conduct the heat away quickly to the neighbouring tissues. This also causes the tissue regions surrounding the calcifications to rise in temperature.

Perhaps surprisingly, the thermal dose volumes in the presence of calcifications are slightly higher than when sonicating the same duration without calcifications. This is due to the geometry of the simulations: the simulations were conducted with a fixed transducer position (i.e., no rotation), and hence the regions `below' and `above' the calcifications obtained some thermal damage due to thermal conduction. This caused the thermal dose volume to grow in the elevation direction (i.e., the z-axis). Without calcifications, the thermal dose volumes in the elevation direction are smaller. In a real treatment situation, however, the rotational movement of the ultrasound probe would treat the regions below and above the calcifications. The results between different patients cannot be directly compared due their different prostate sizes as well as different sizes and locations of the calcifications.


\subsection*{Study with artificial calcifications}

\begin{figure}[b!]
\centering
\subfigure[]{\includegraphics[height=4cm]{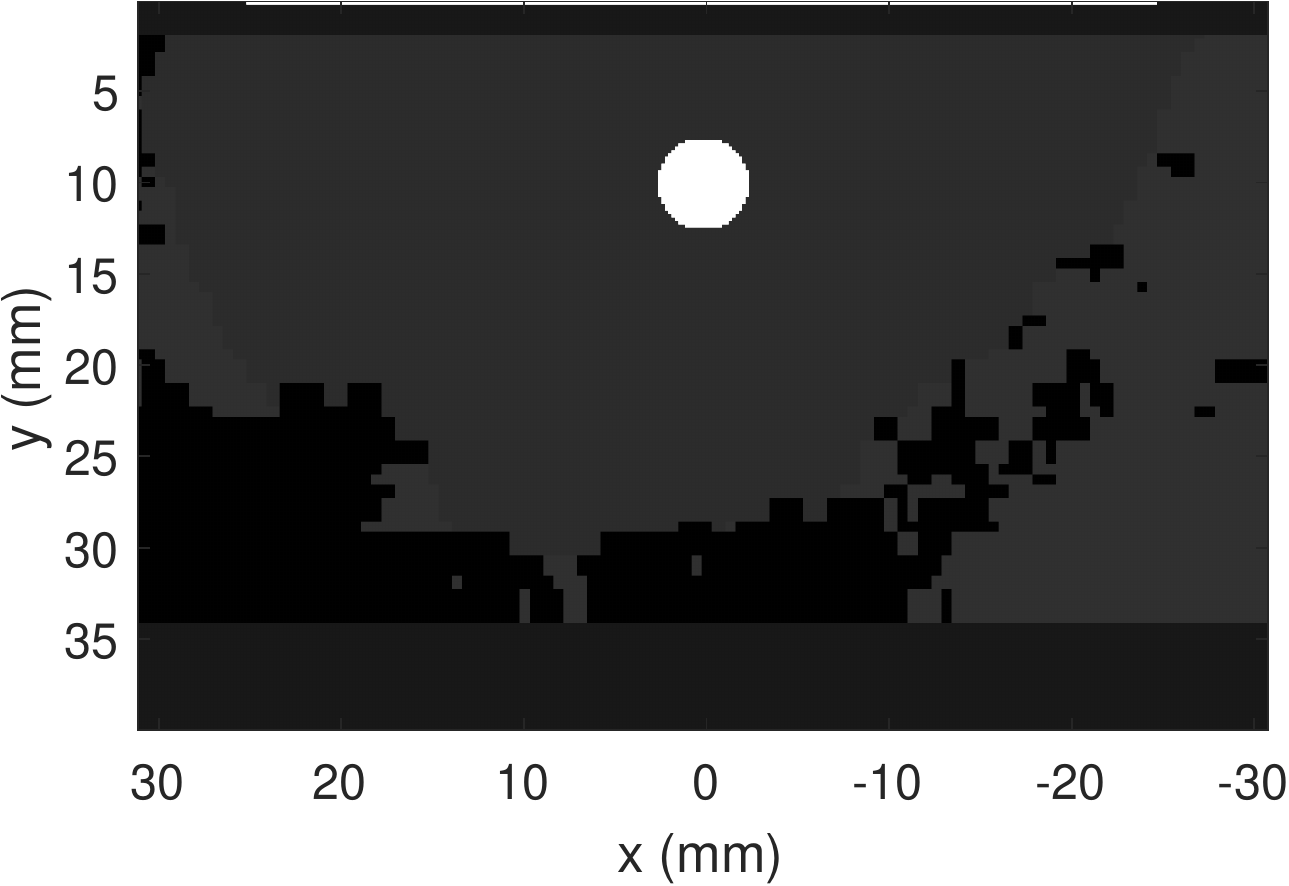}}
\subfigure[]{\includegraphics[height=4cm]{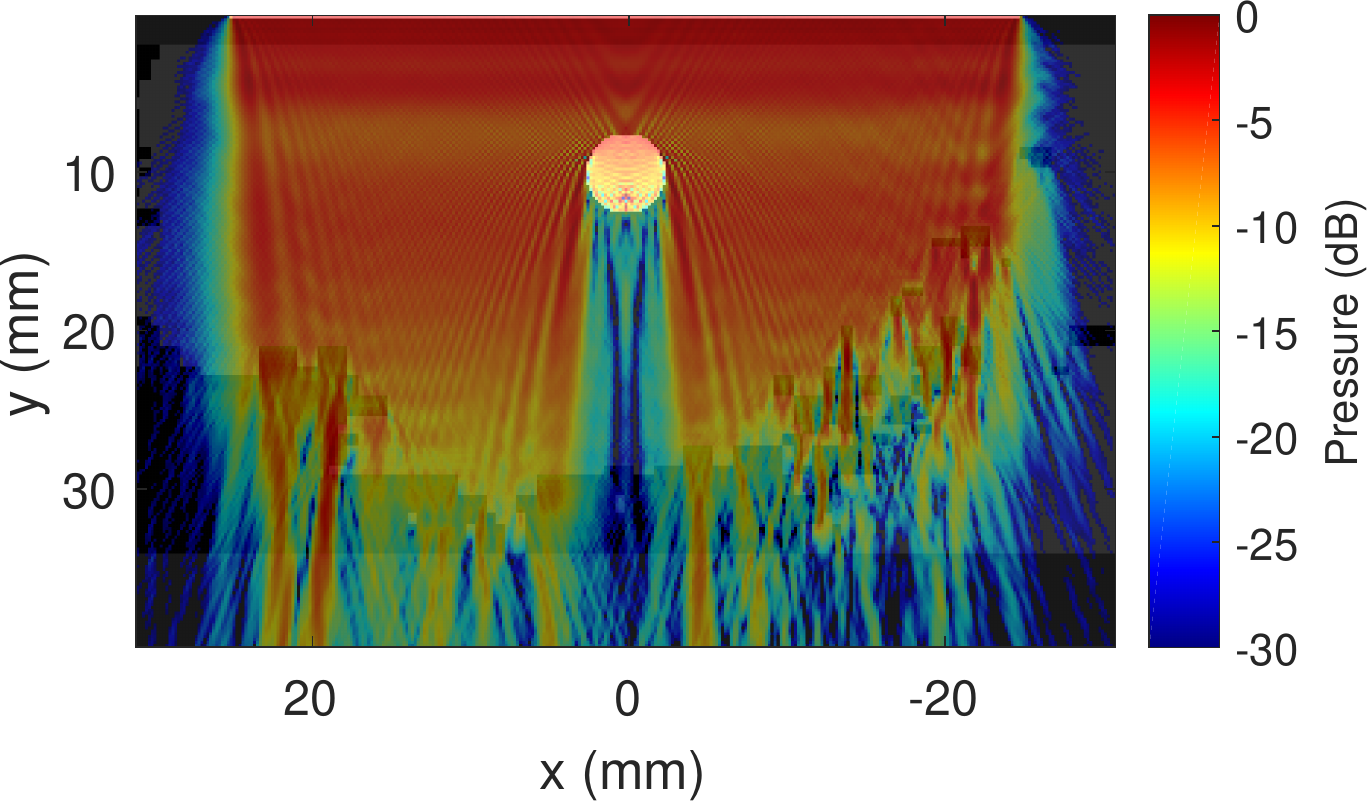}}
\\
\subfigure[]{\includegraphics[height=4cm]{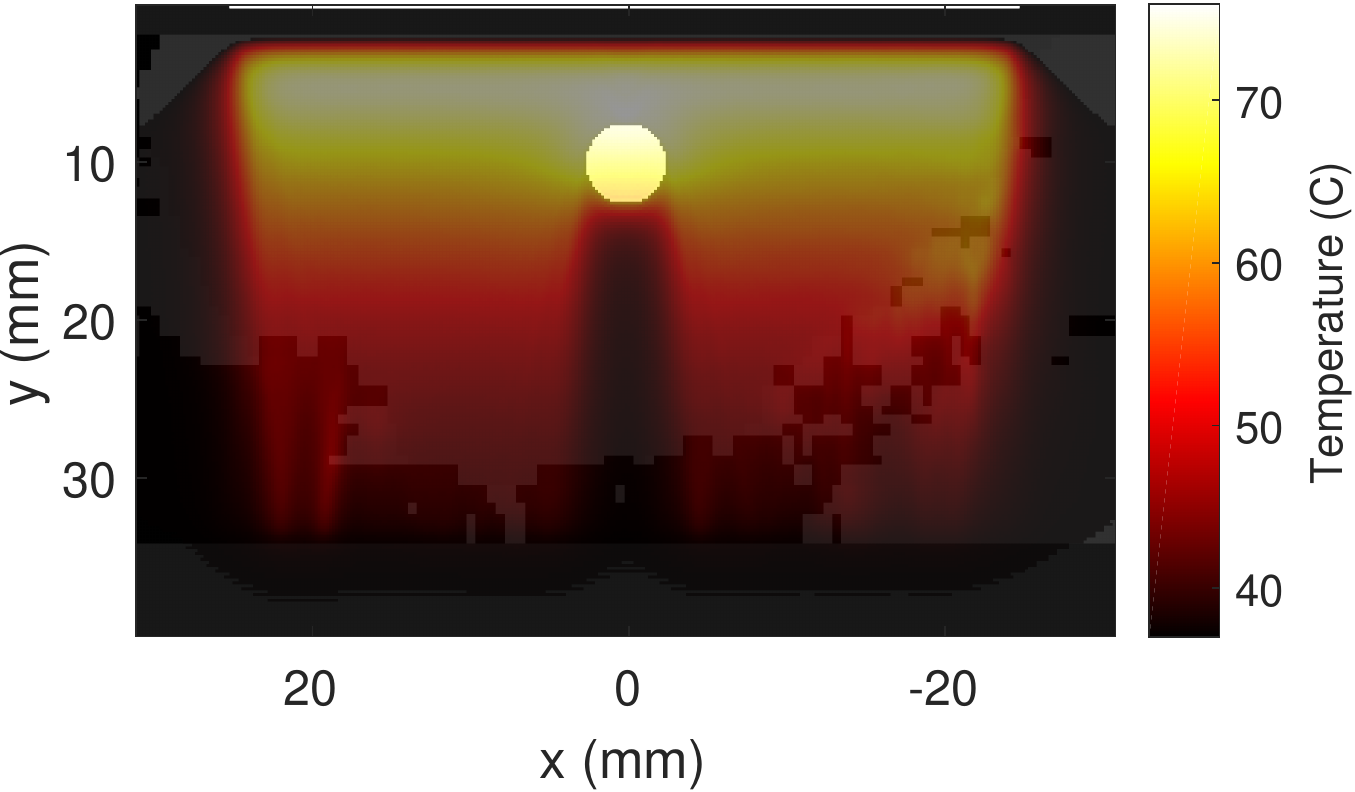}}
\subfigure[]{\includegraphics[height=4cm]{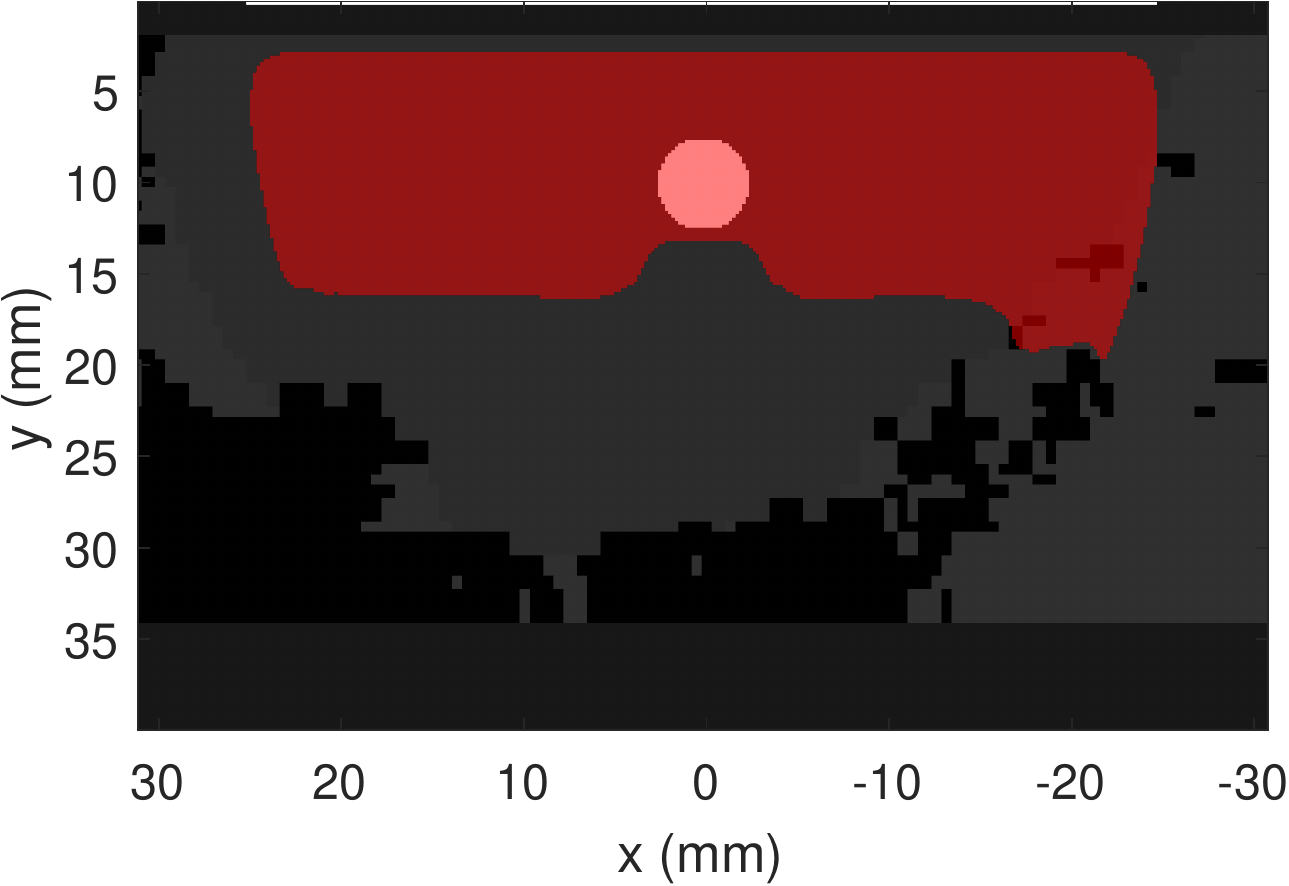}}
\caption{Segmented computed tomography (CT) slice of the half of the prostate (grey) with an artificial spherical calcification (white). The tissue areas surrounding the prostate were segmented as fat (black) and muscle (light grey). The calcification had a diameter of 5~mm and its centre was located 10~mm from the transducer along y-axis. Simulated (b) ultrasound, (c) temperature and (d) 240 cumulative equivalent minutes (CEM) thermal dose fields at the end of the sonication.}
\label{fig:artificial_planes}
\end{figure}

The study with artificial calcifications was conducted by using spherical calcifications of different sizes and locations inside the prostate. Furthermore, different sonication durations and frequencies were used in order to examine their effect on the temperature evolution and thermal dose areas. The thermal dose areas were quantified instead of the volumes, because only the area is clinically significant due to the rotational movement of the transducer. The volume surrounding the calcification in the z-direction (elevation) will be ablated by the rotation of the transducer regardless of the presence of a calcification. The simulation geometry with a 5~mm diameter calcification located at 10~mm distance from the centre of the transducer is shown in Figure~\ref{fig:artificial_planes}(a) together with simulated (b) ultrasound, (c) temperature and (d) 240~CEM thermal dose fields.

\begin{figure}[htbp!]
\vspace{-0.5cm}
\centering
\subfigure[]{\includegraphics[height=2.8cm]{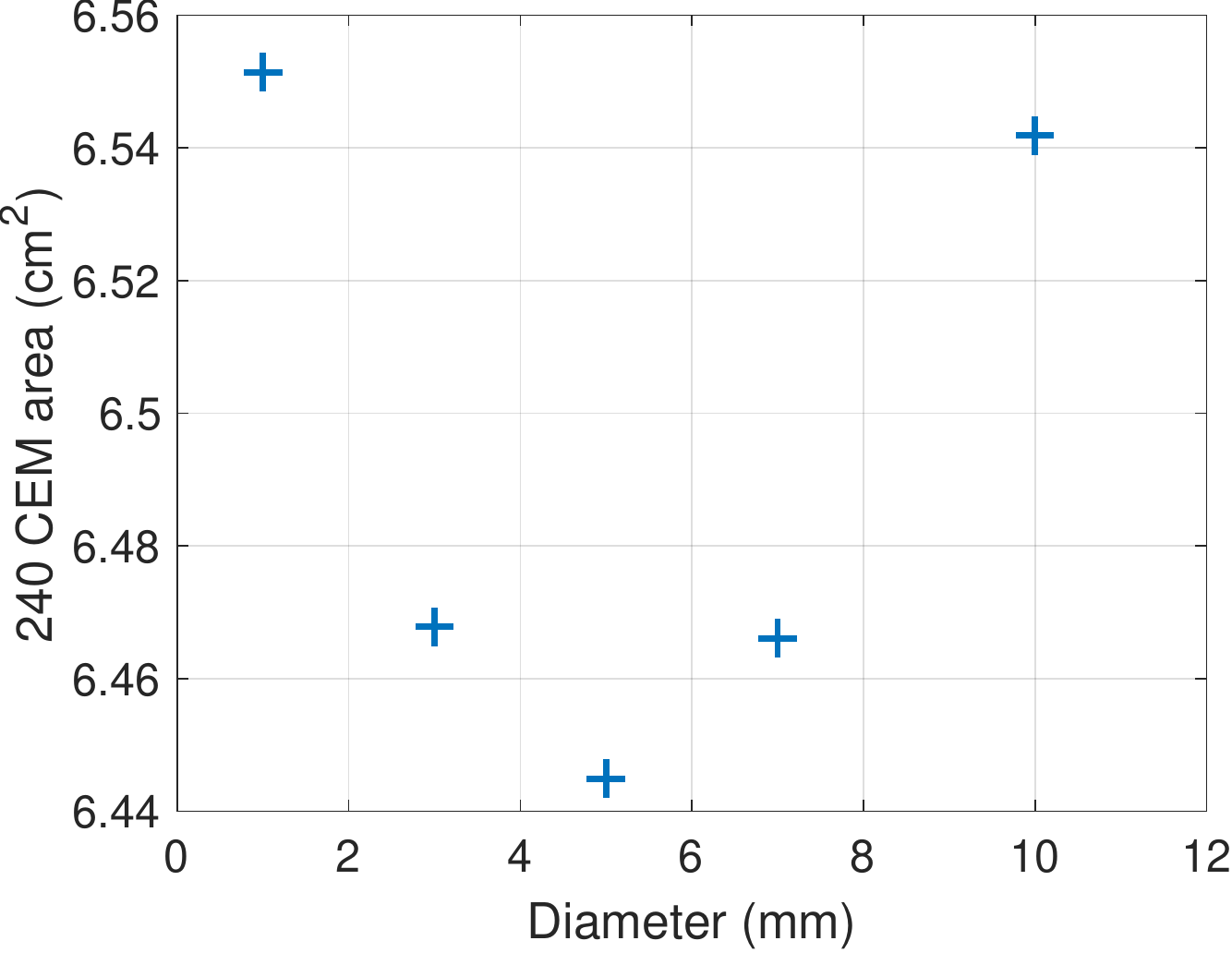}}
\subfigure[]{\includegraphics[height=2.8cm]{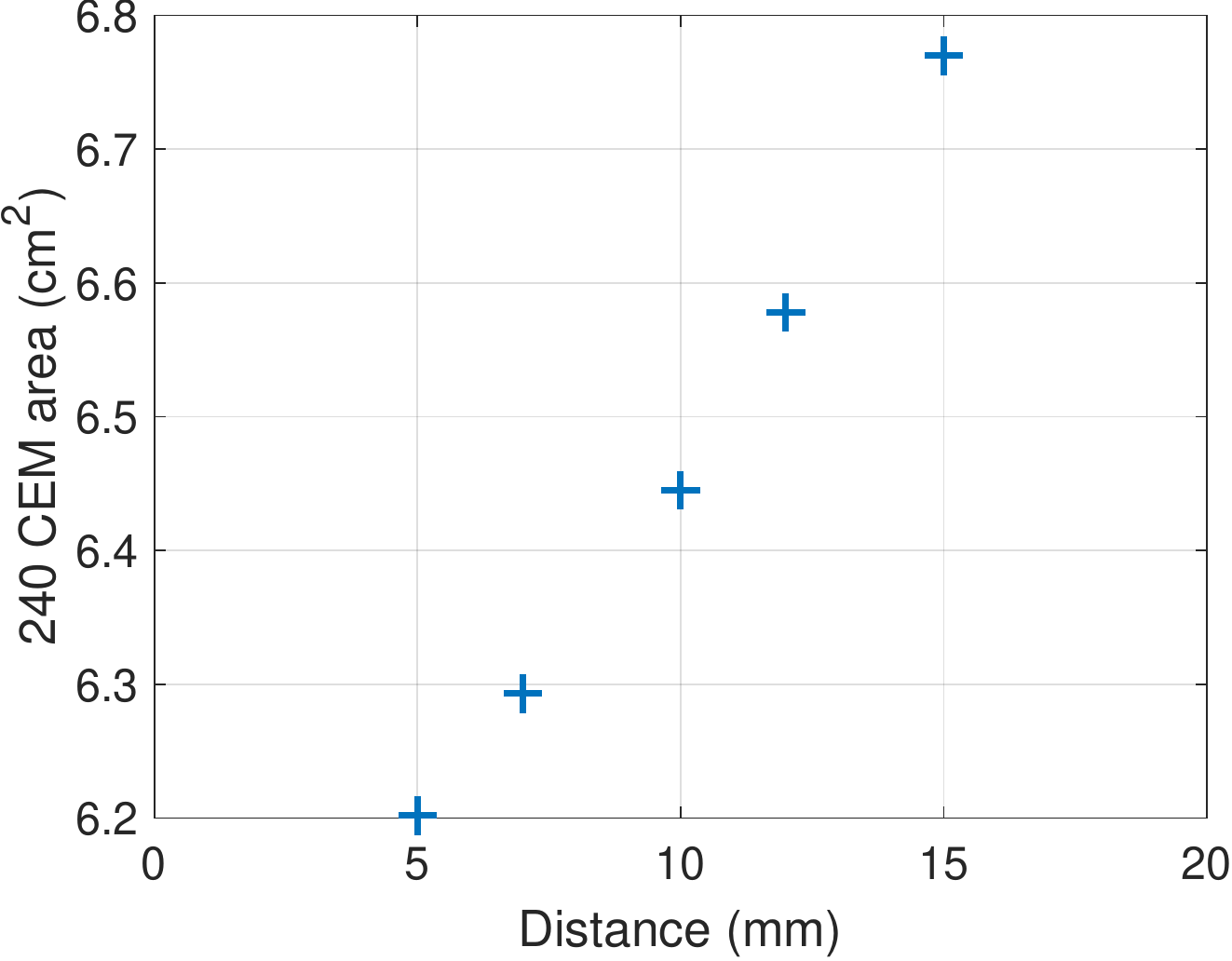}}
\\
\subfigure[]{\includegraphics[height=2.8cm]{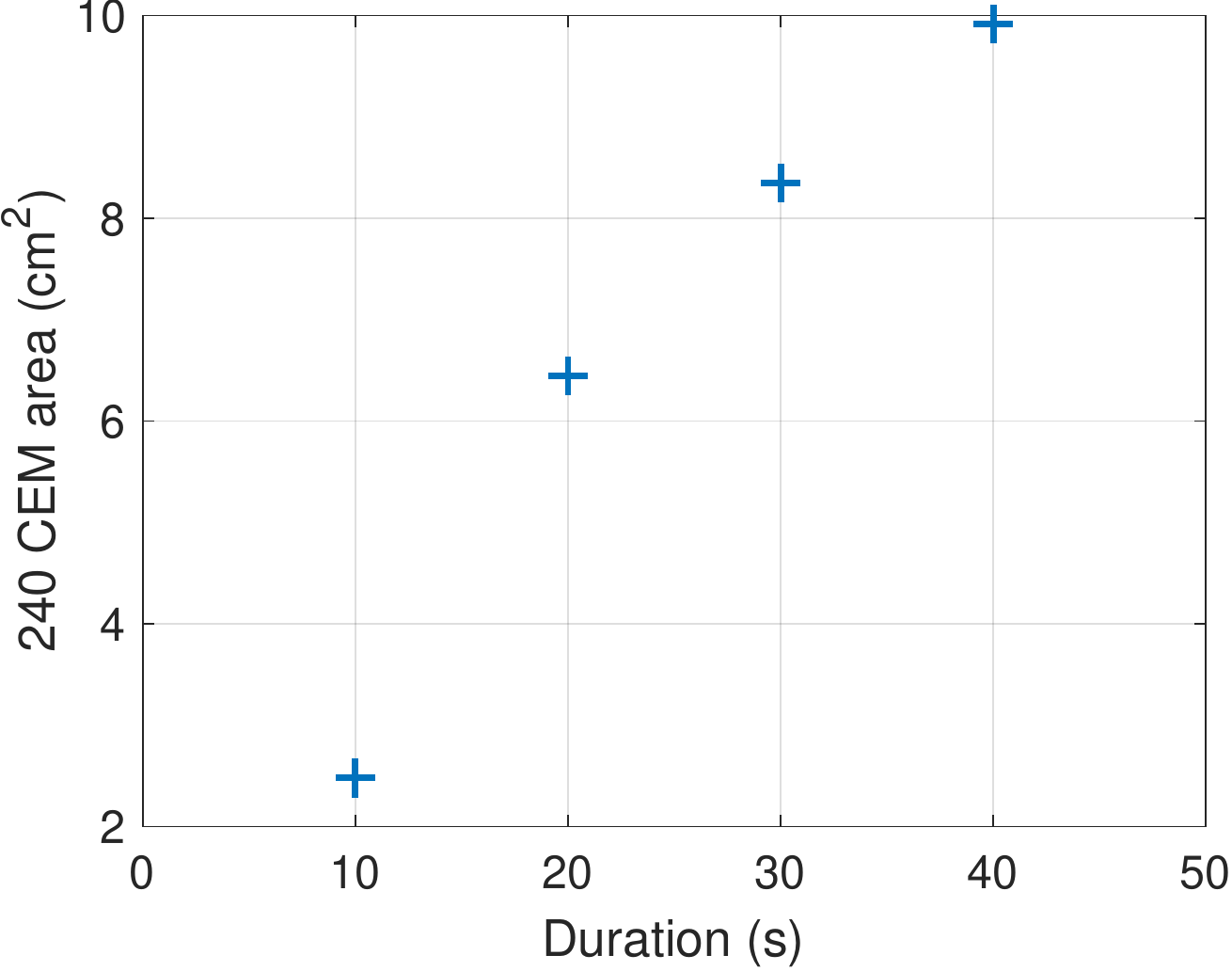}}
\subfigure[]{\includegraphics[height=2.8cm]{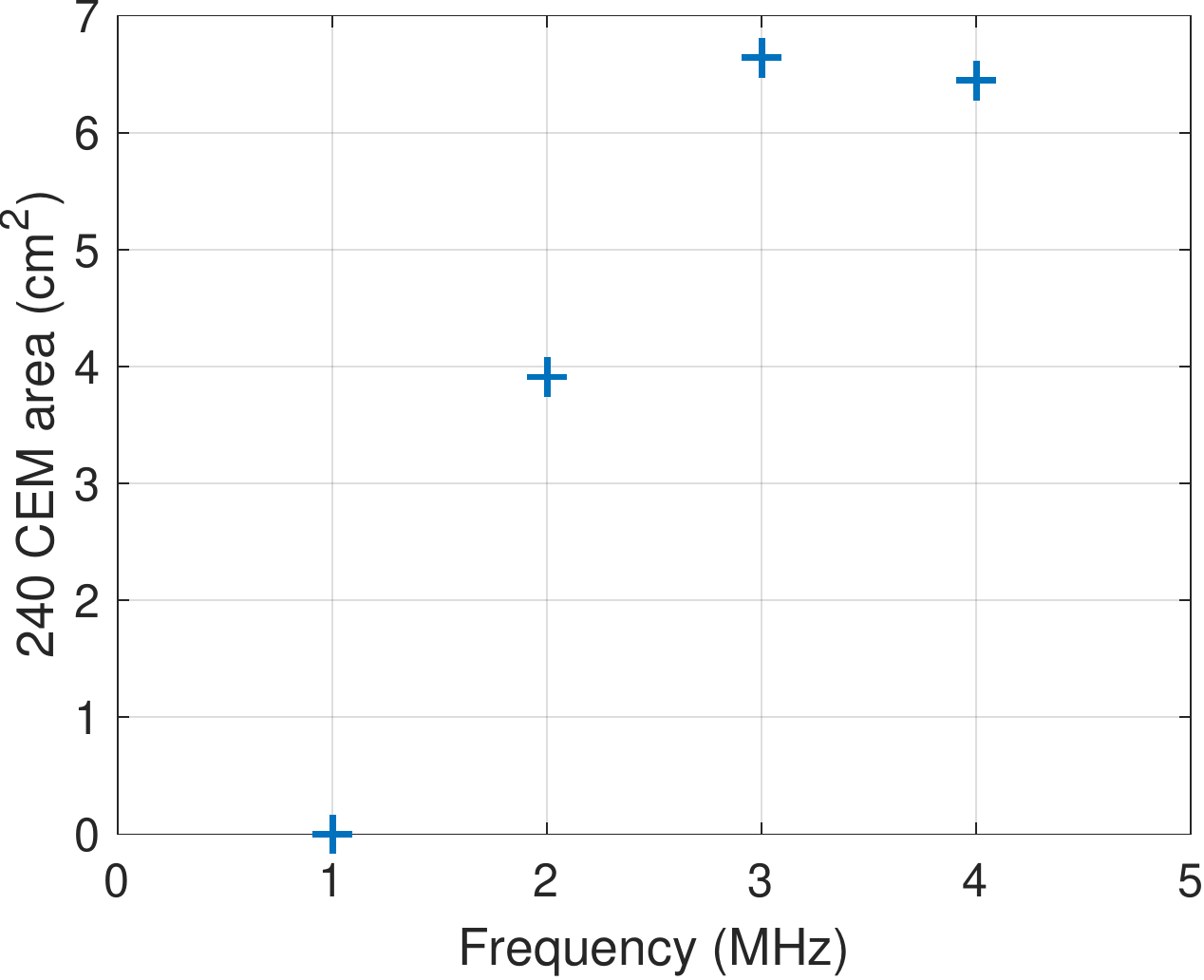}}
\\
\subfigure[]{\includegraphics[height=2.8cm]{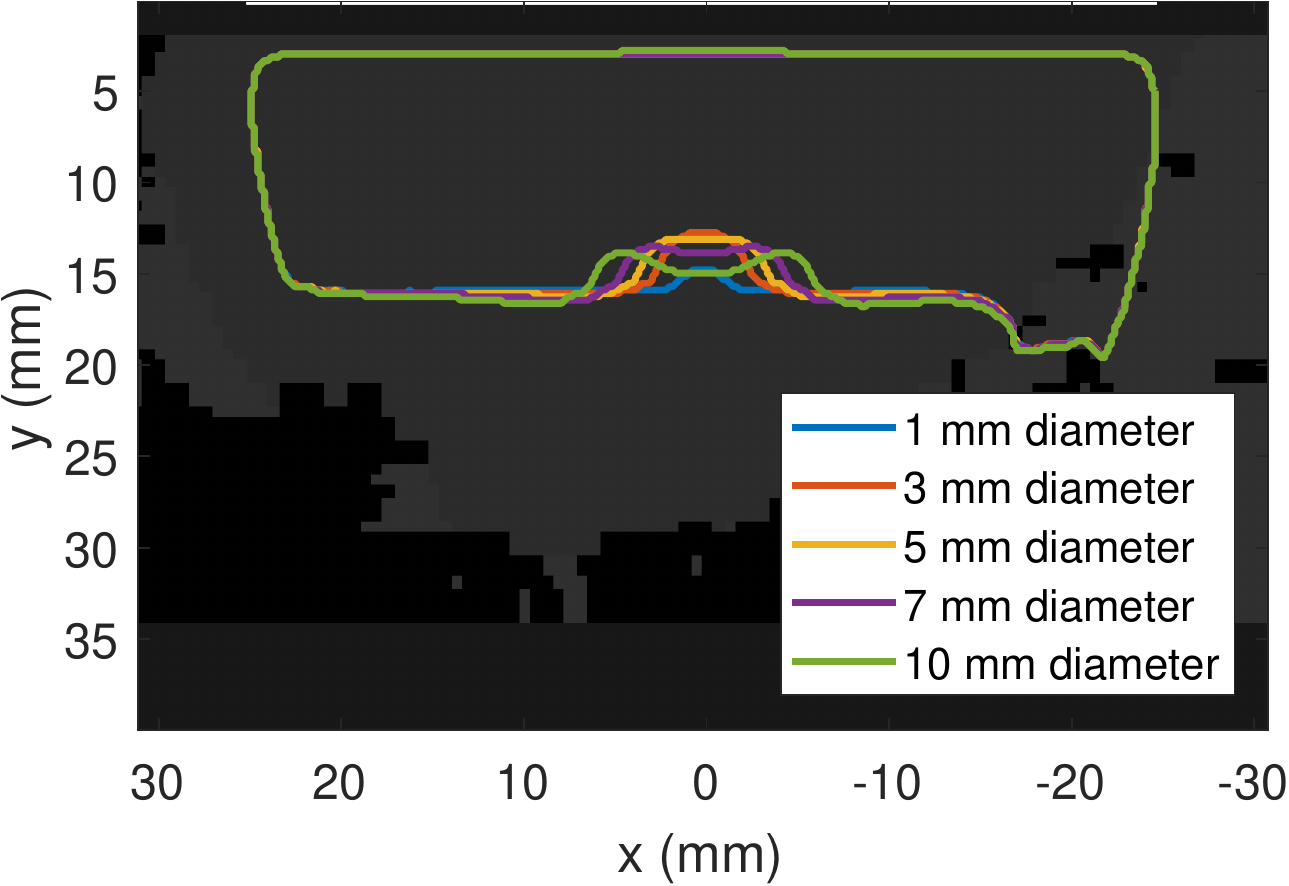}}
\subfigure[]{\includegraphics[height=2.8cm]{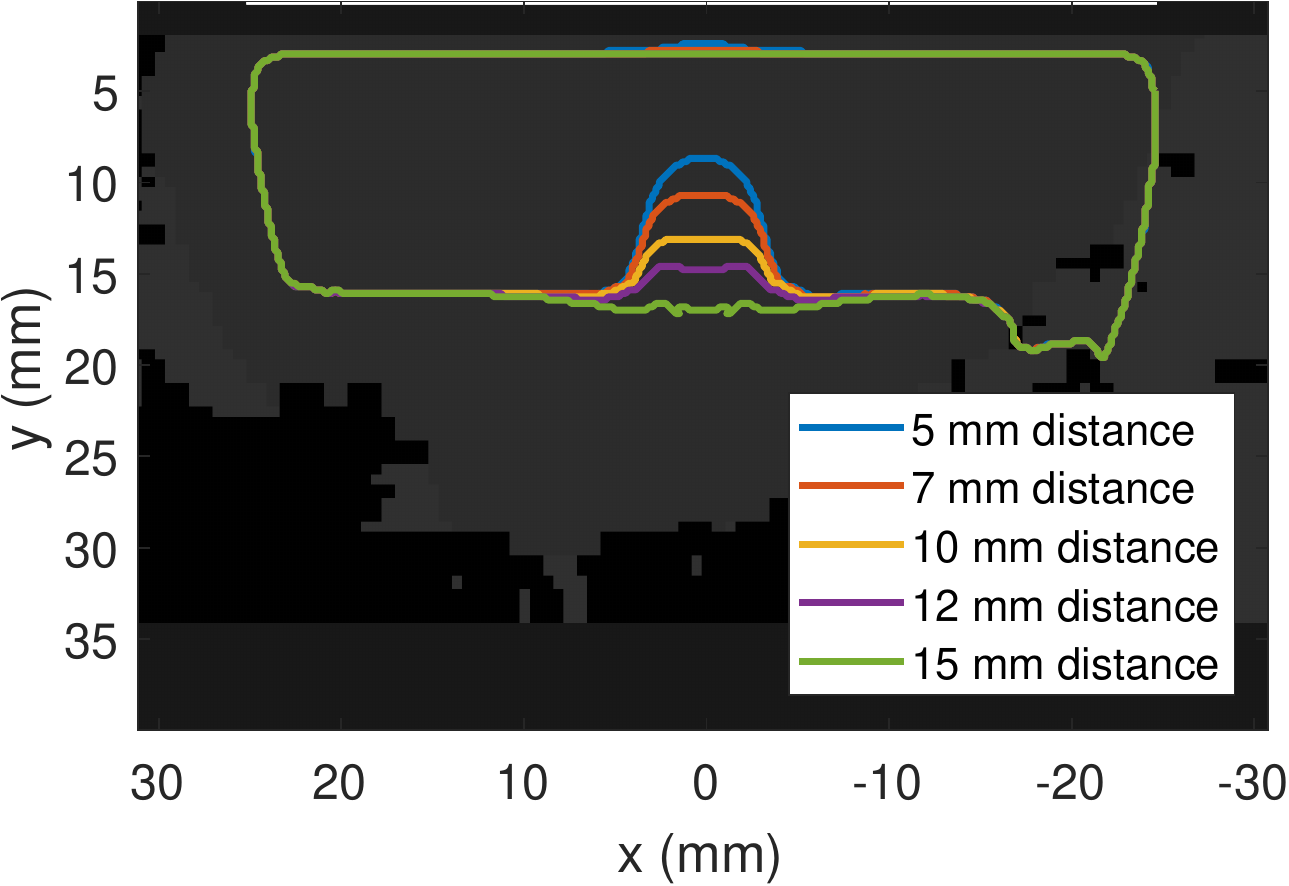}}
\\
\subfigure[]{\includegraphics[height=2.8cm]{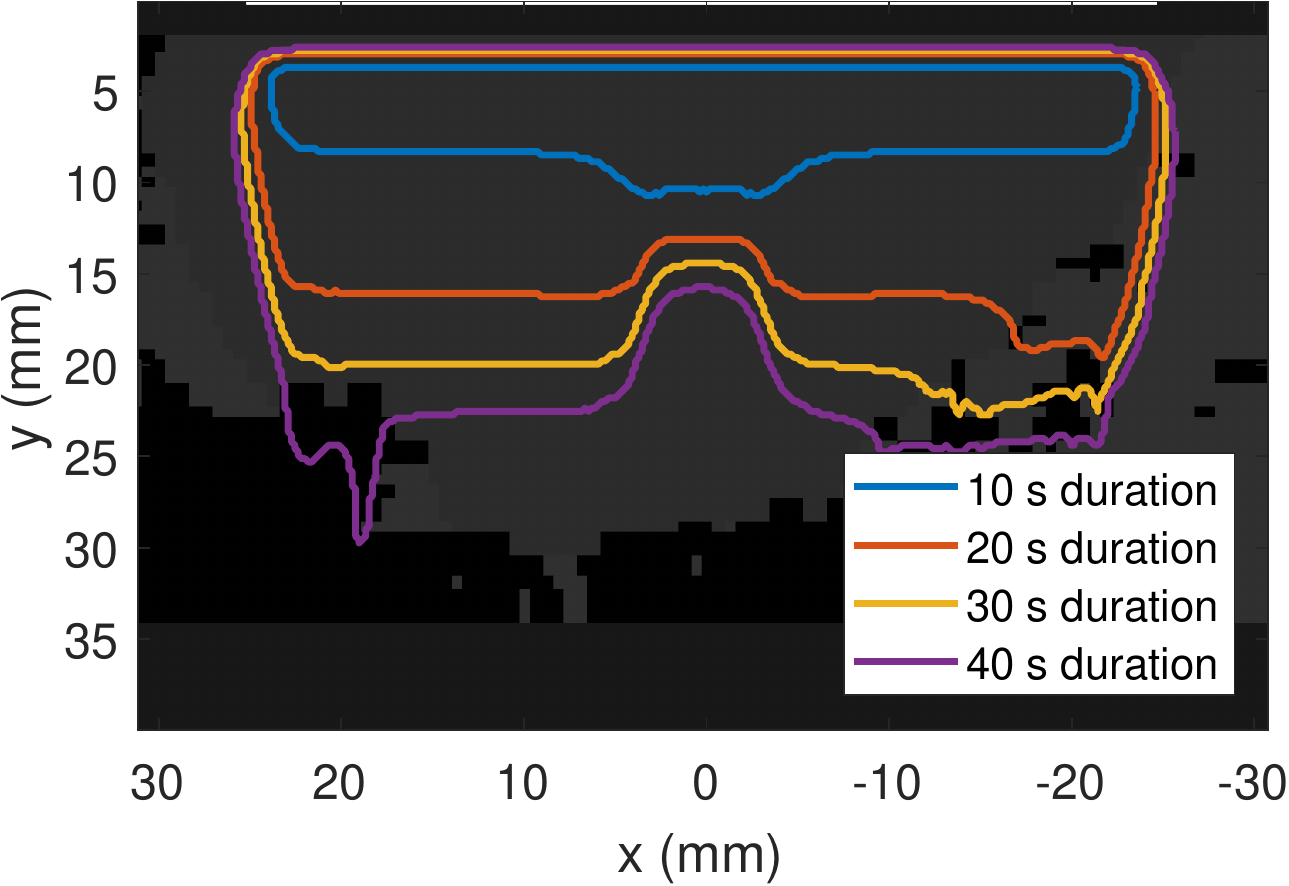}}
\subfigure[]{\includegraphics[height=2.8cm]{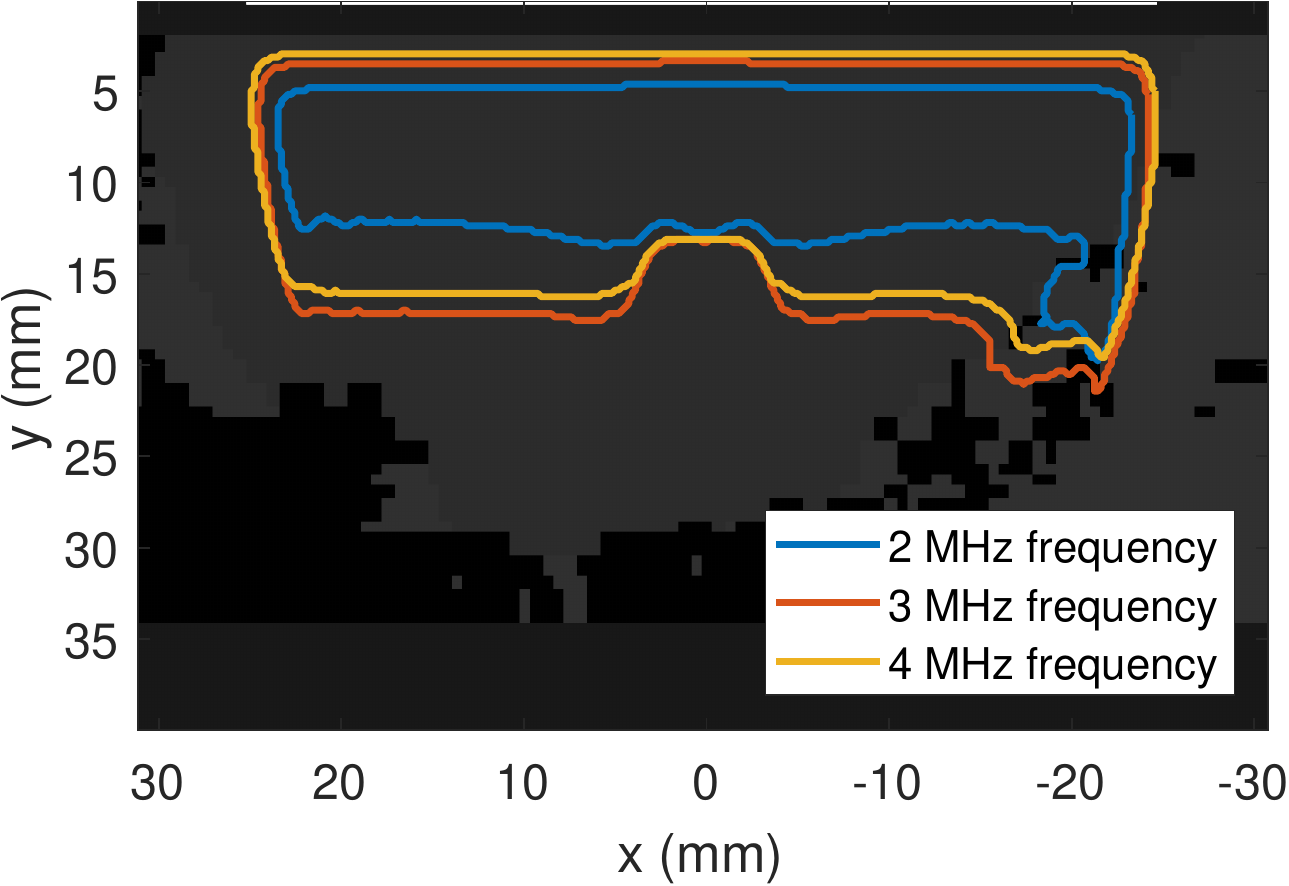}}
\caption{The dependence of 240~CEM thermal dose area upon (a) the diameter of the calcification, (b) the distance of the calcification centre from the transducer, (c) sonication duration and (d) sonication frequency. (e)-(h) Contour lines showing the boundaries of the same thermal dose areas on an axial plane crossing the centre of the transducer, respectively. The sonication duration was 20 seconds followed by a 40-second cooling time in all cases except for (c) and (g) where the sonication duration was varied.}
\label{fig:artificial_dependence}
\end{figure}

Figure~\ref{fig:artificial_dependence}(a) shows the evolution of the thermal dose area when the diameter of the spherical calcification was changed while its central distance from the transducer was kept constant at 10~mm. In Figure~\ref{fig:artificial_dependence}(e), the same areas are shown with 2-D contours on the axial plane cutting the centre of the transducer. The thermal dose areas reduce with the diameter of the calcification up to 5~mm after which the area starts to grow again. This is because at this point the diameter of the calcification becomes so large that the thermal damage extends beyond the areas of smaller calcifications.

Figure~\ref{fig:artificial_dependence}(b) shows the dependence of thermal dose area upon the distance of the central point of the calcification from the centre of the transducer while the diameter was kept constant at 5~mm. Figure~\ref{fig:artificial_dependence}(f) shows the thermal dose area boundaries on the axial plane. It should be noted that the distance of the outer surface of the calcification in Figure~\ref{fig:artificial_dependence} is actually 2.5~mm closer than the indicated central point. This means that the outer surface of the nearest calcification in the figure is actually only 2.5~mm away from the transducer and 0.5~mm away from the ultrasound probe. 

In Figure~\ref{fig:artificial_dependence}(b), the thermal dose area grows as the distance of the calcification from the transducer becomes larger. There exists an almost linear dependence of the thermal dose area on the distance. This is somewhat expected because the calcification almost completely blocks the ultrasound field from penetrating to the rear region thus diminishing the heating effect. The closer the calcification is located to the transducer, the larger the region behind it that remains untreated. When the calcification is located very close to the transducer, some thermal damage is induced to the tissue near the urethra (see the upper blue and red boundaries in Figure~\ref{fig:artificial_dependence}(f)). The same damage is not present in the sonications with calcifications at further distances (i.e, 10~mm and above), which indicates that calcifications located very close to the transducer might cause thermal damage to the urethra.

The dependence of thermal dose area upon the sonication duration with fixed calcification diameter (5~mm) and location from the transducer (10~mm) is shown in Figure~\ref{fig:artificial_dependence}(c). The axial planes of the same areas are shown with contours in Figure~\ref{fig:artificial_dependence}(g).
The thermal dose area grows with sonication duration, but the growth rate slows slightly with longer sonication durations. The axial contours show that the treated region behind the calcification grows slightly with sonication duration, but the growth speed is still slower than the region boundary without an obstructing calcification. This indicates that simply increasing the sonication duration in order to ablate the region behind the calcifications is not feasible since this could cause damage to adjacent healthy tissue regions. It can also be seen that with a 10-second sonication duration, the damaged tissue region behind the calcification actually extends further than it otherwise would without the calcification in place. This is caused by the dissipation of heat energy from the calcification to the surrounding tissue regions due to its high thermal conductivity.

Another method to treat the tissue regions behind the calcifications could possibly be the usage of lower ultrasound frequencies due to the lower attenuation, and hence, higher penetration depth. Figure~\ref{fig:artificial_dependence}(d) illustrates the dependence of thermal dose area upon sonication frequency for a fixed calcification diameter (5~mm) and distance from the transducer (10~mm). The axial contour profiles of the same areas are shown in Figure~\ref{fig:artificial_dependence}(h). The thermal dose areas grow with sonication frequency up to 3~MHz after which the area becomes smaller again. This is an interesting observation because the heating rate is directly dependent on frequency, and therefore, one would expect the area to grow with higher frequencies. However, higher frequencies are also more strongly attenuated in the tissue. Thus, higher frequencies can be seen to heat more in the near field while lower frequencies extend further.

When using a frequency of 1~MHz, the 20-second sonication duration was not long enough to cause any thermal damage above 240~CEM. The hypothesis of better penetration of ultrasound to the region behind the calcification with lower frequencies also does not hold true: the untreated regions behind the calcification between 2, 3 and 4~MHz frequencies have almost the same depth. This is due to the fact that large acoustic impedance mismatch still exists between the prostate and the calcification although the effect of attenuation is smaller with lower frequencies.

\begin{figure}[htbp!]
\vspace{-0.5cm}
\centering
\subfigure[]{\includegraphics[height=5.1cm]{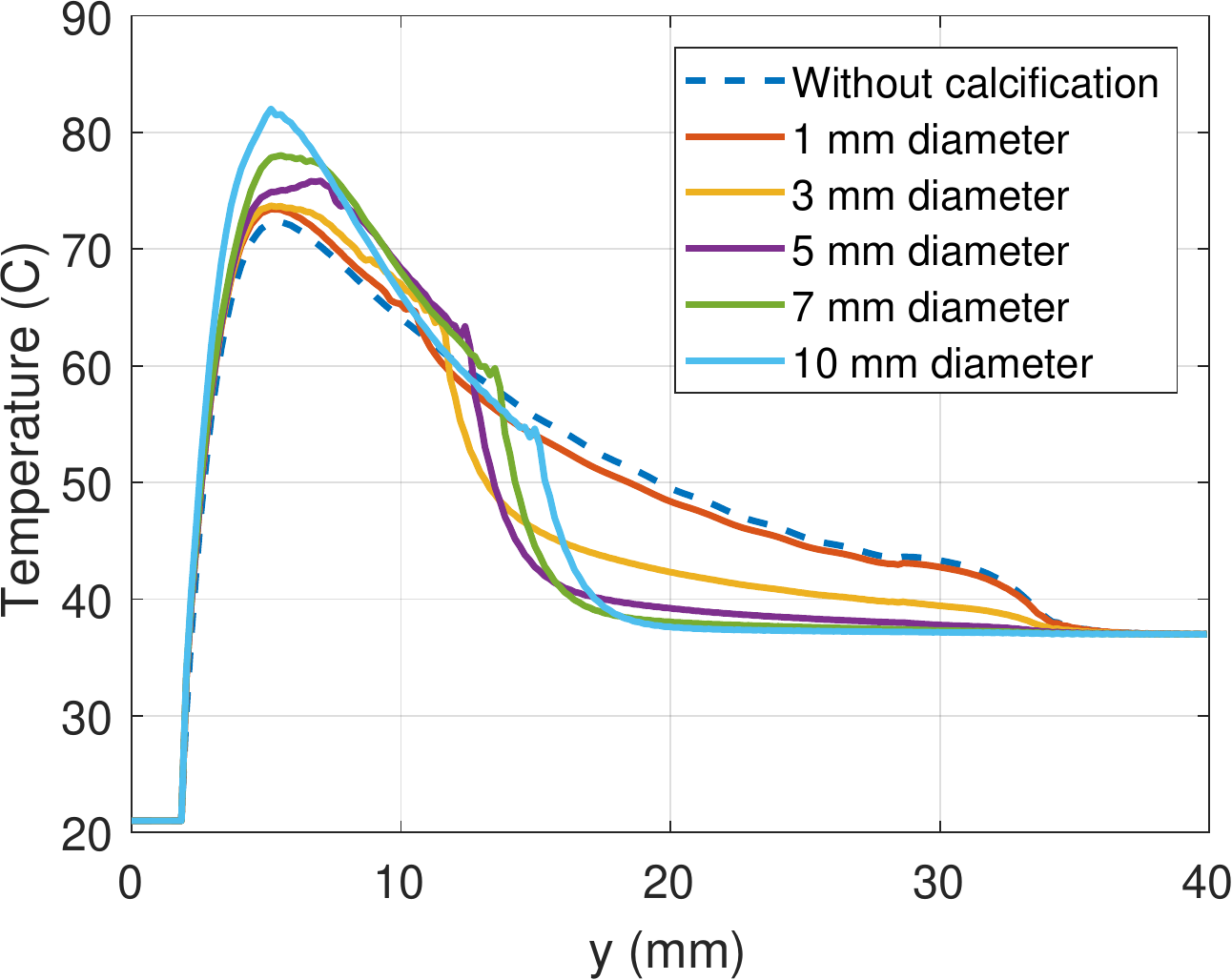}}
\subfigure[]{\includegraphics[height=5.1cm]{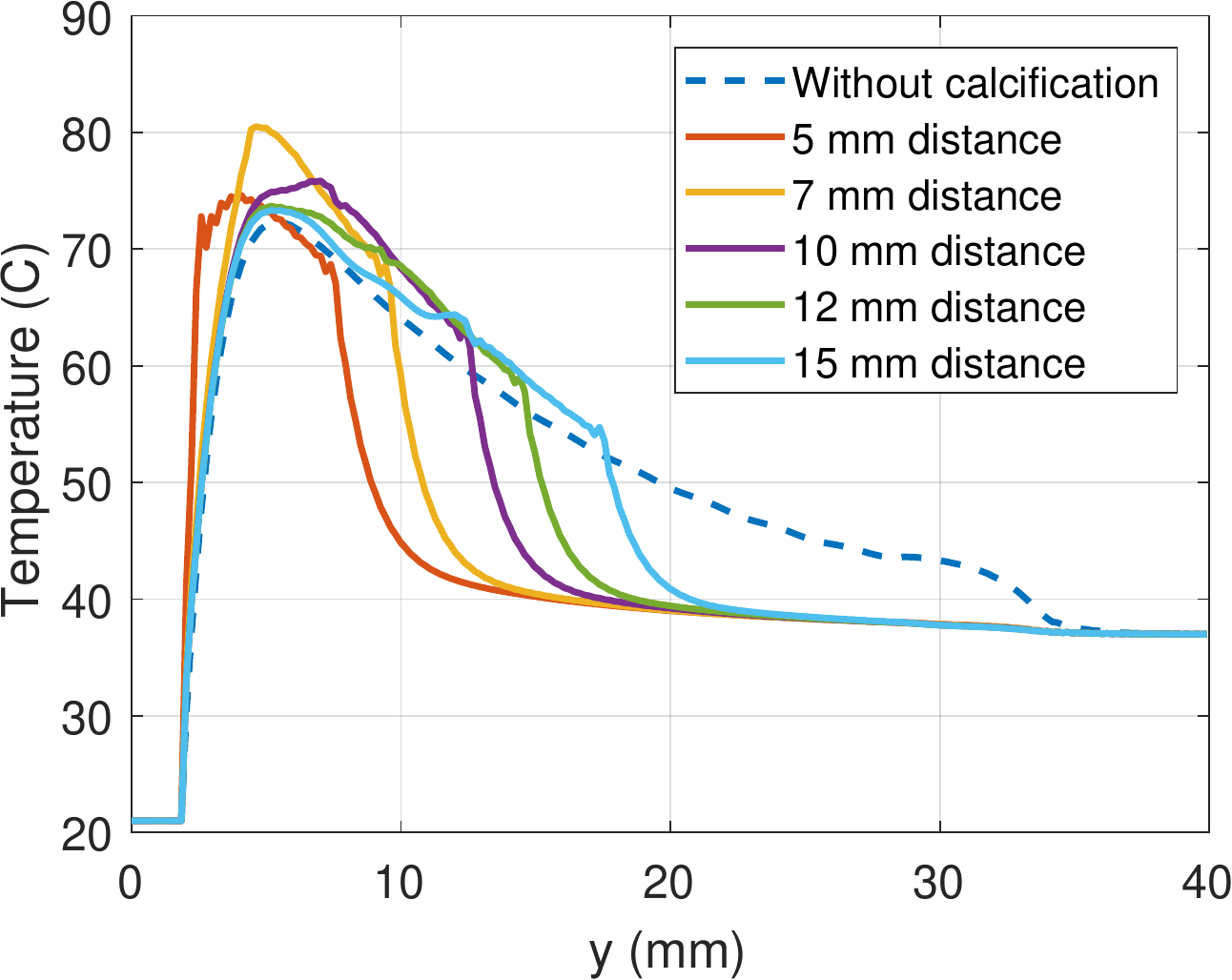}}
\\
\subfigure[]{\includegraphics[height=5.1cm]{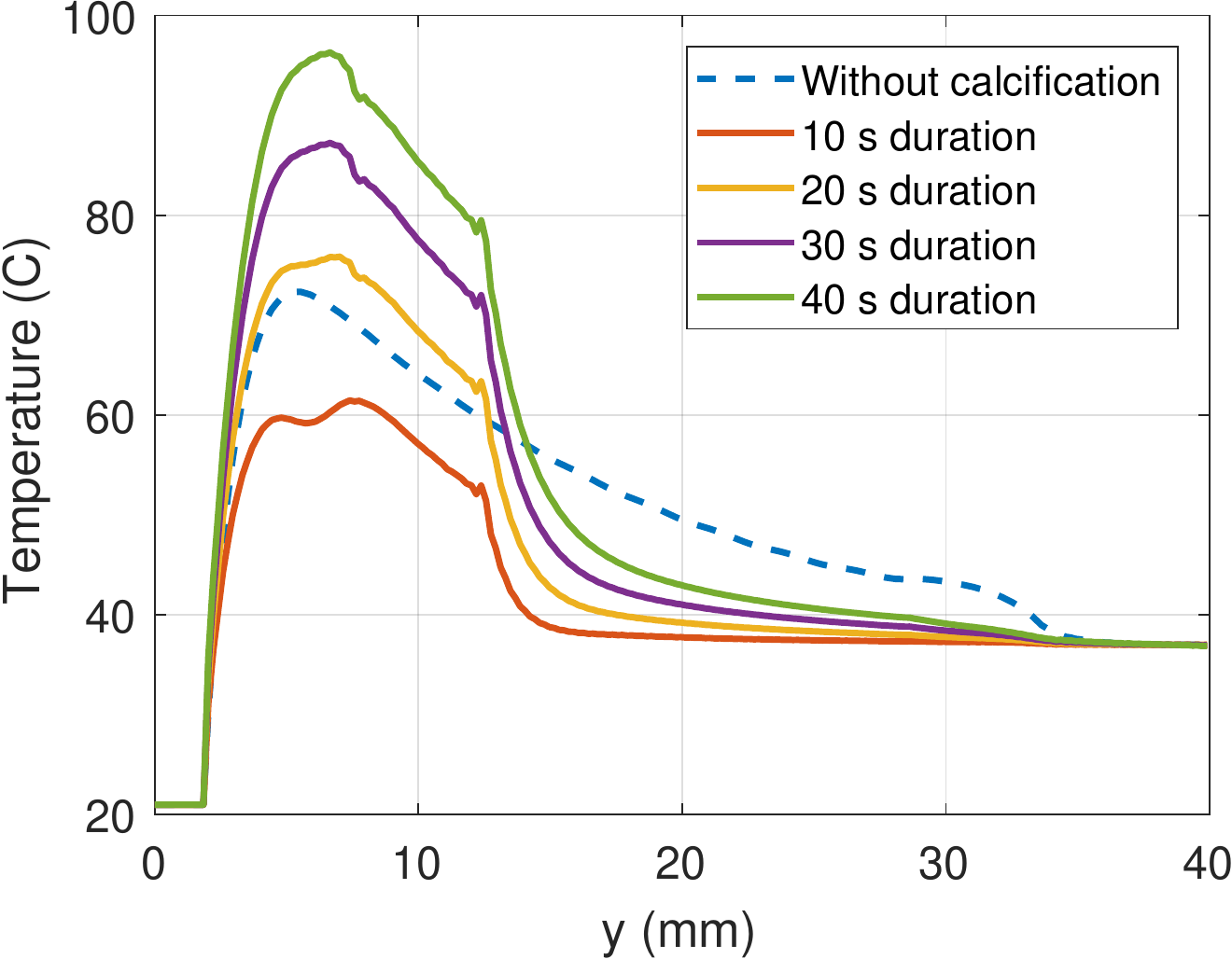}}
\subfigure[]{\includegraphics[height=5.1cm]{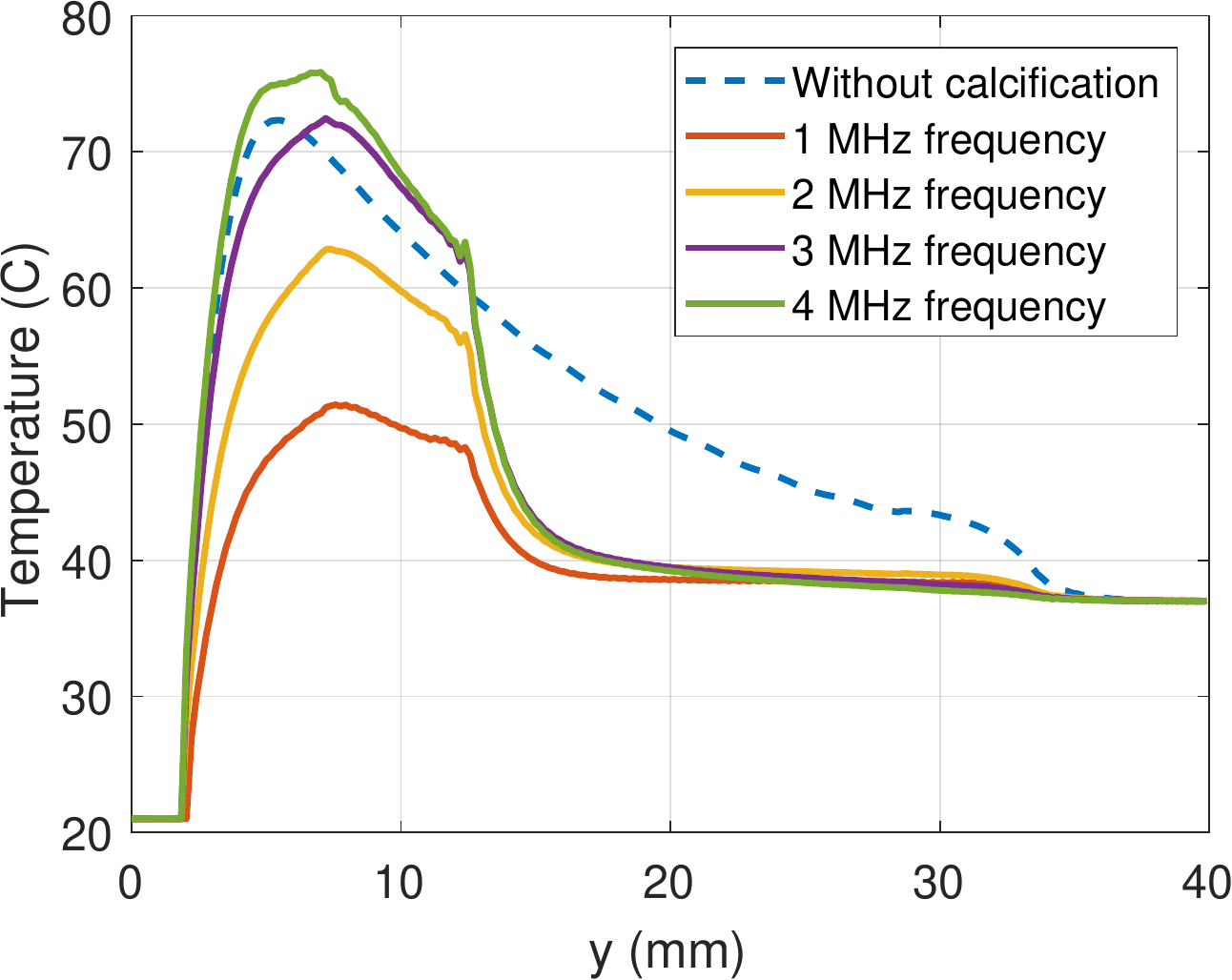}}
\caption{Axial temperature profiles from the centre of the transducer (the origin) along the y-axis with (a) a spherical calcification with different diameters with the centre located at 10~mm distance from the transducer, (b) a spherical calcification with a fixed 5~mm diameter with the centre located at different distances from the transducer, (c)-(d) a spherical calcification with a fixed diameter (5~mm) and distance (10~mm) from the centre of the transducer with different sonication durations and frequencies, respectively. A simulation without a calcification (4~MHz, 20~s) is also shown in each figure for reference.}
\label{fig:artificial_axial}
\end{figure}

The axial temperature profiles along the central axis of the transducer at the end of the 20-second sonication are shown in Figure~\ref{fig:artificial_axial} for all the cases described above. Figure~\ref{fig:artificial_axial}(a) shows the axial temperature profiles with different calcification diameters located 10~mm from the transducer. The temperature profile without the calcification is also shown for reference. In all cases, the temperature sharply increases towards the calcification and peaks at around 5~mm. Slightly higher temperatures in the proximity of the transducer are seen when calcifications are present. After the peak, an almost linear temperature decrease is seen inside the calcification as the heat energy dissipates from the front of the calcification towards the cooler rear side and further into the prostate tissue. When again entering to the region of the prostate tissue, a sharper drop in the temperature is seen. This is most likely caused by perfusion carrying the heat energy away, which is not the case inside the calcification. The region behind the calcification also stays cooler due to smaller heating rate of the ultrasound field.

Figure~\ref{fig:artificial_axial}(b) shows the axial temperature profiles with a fixed diameter (5~mm) calcification located at different distances from the transducer. As the calcification is moved further away from the transducer, the location of the linear temperature decrease and the sharper temperature drops also shift. When the calcification is very close to the transducer, the increase in temperature in the beginning is more rapid, which indicates some of the heat energy is accumulating between the transducer and the calcification despite of the constant temperature ultrasound probe acting as a heat sink. During the clinical treatments, this could possibly cause thermal damage to the urethra.

Figure~\ref{fig:artificial_axial}(c) shows the axial temperature profiles with different sonication durations while the diameter (5~mm) and the position (10~mm) of the calcification was kept constant. The trends are similar to previous figures where linear temperature decreases are seen inside the calcifications followed by a steeper drop when entering the region of the prostate at y = 12.5~mm. The longer the sonication duration, the higher the temperature elevation both in front of and behind the calcifications. The temperature curves also seem to peak right in front of the calcifications.

Finally, the effect of different sonication frequencies on the axial temperature profiles with a fixed calcification diameter (5~mm) and location (10~mm) is shown in Figure~\ref{fig:artificial_axial}(d). Since the heating rate is dependent on the frequency, sonications with higher frequencies exhibit higher temperatures when the sonication duration and power are kept constant. Again, similar characteristics are seen in the profiles as observed before with a linear decrease in temperature inside the calcification and a more pronounced drop in temperature when entering the prostate tissue from the rear side of the calcification. The temperatures behind the calcification are not much larger with higher frequencies, which further indicates that they cannot penetrate effectively through the interfaces of high acoustic impedance mismatch.

\subsection*{Sonication strategies}

Encouraged by the results from the sonications with artificial calcifications, different sonication strategies were tested on naturally occurring calcifications in different patients. The results from the artificial study showed that the best strategy is to vary the sonication duration in order to ablate the tissue regions behind the calcifications.

\begin{figure}[b!]
\centering
\subfigure[]{\includegraphics[height=3cm]{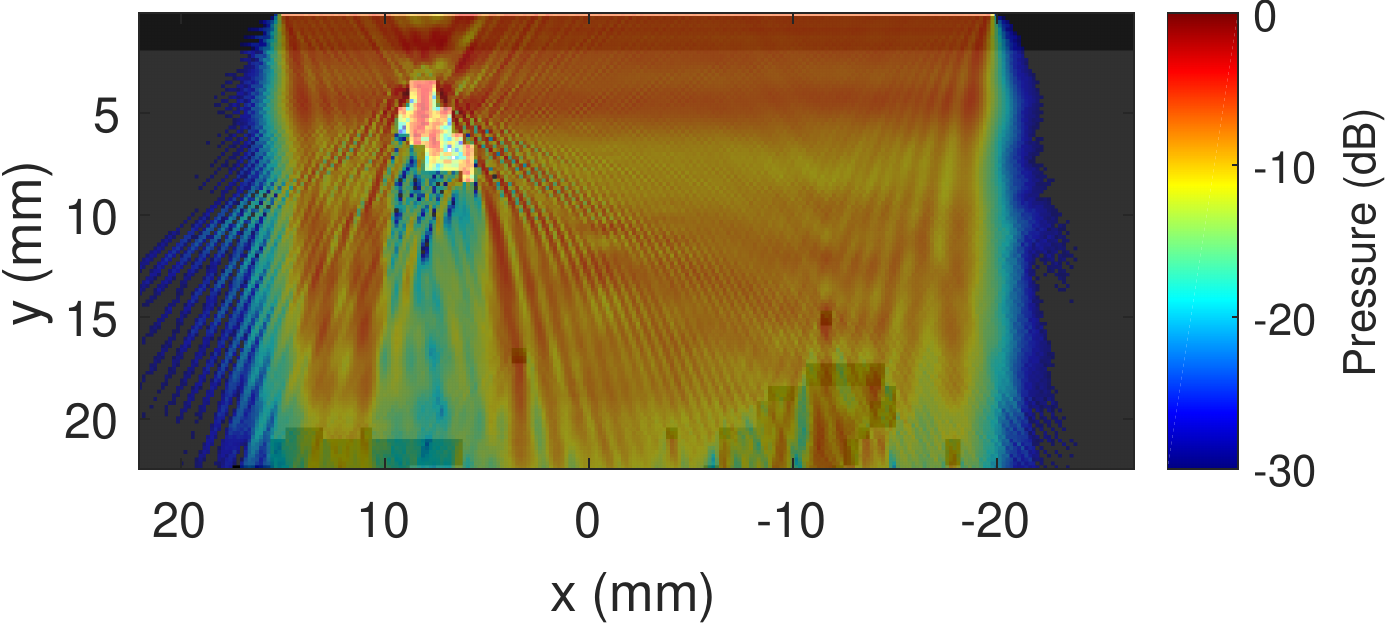}}
\subfigure[]{\includegraphics[height=3cm]{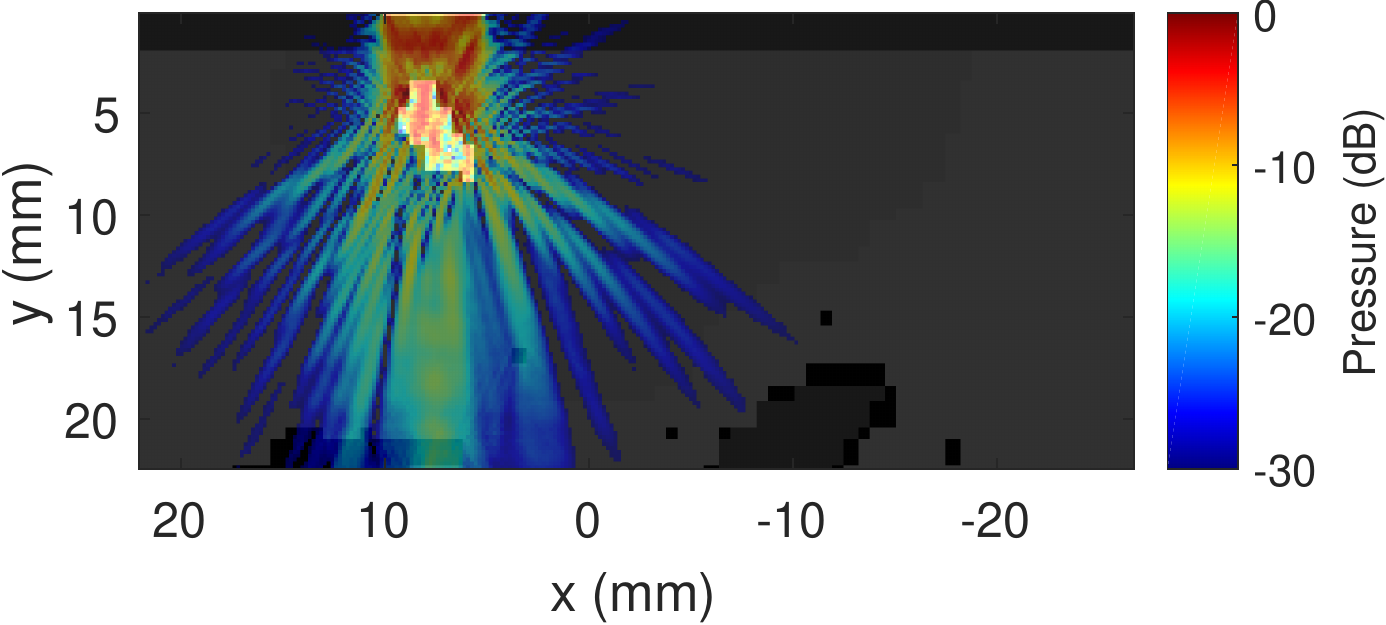}}
\subfigure[]{\includegraphics[height=3cm]{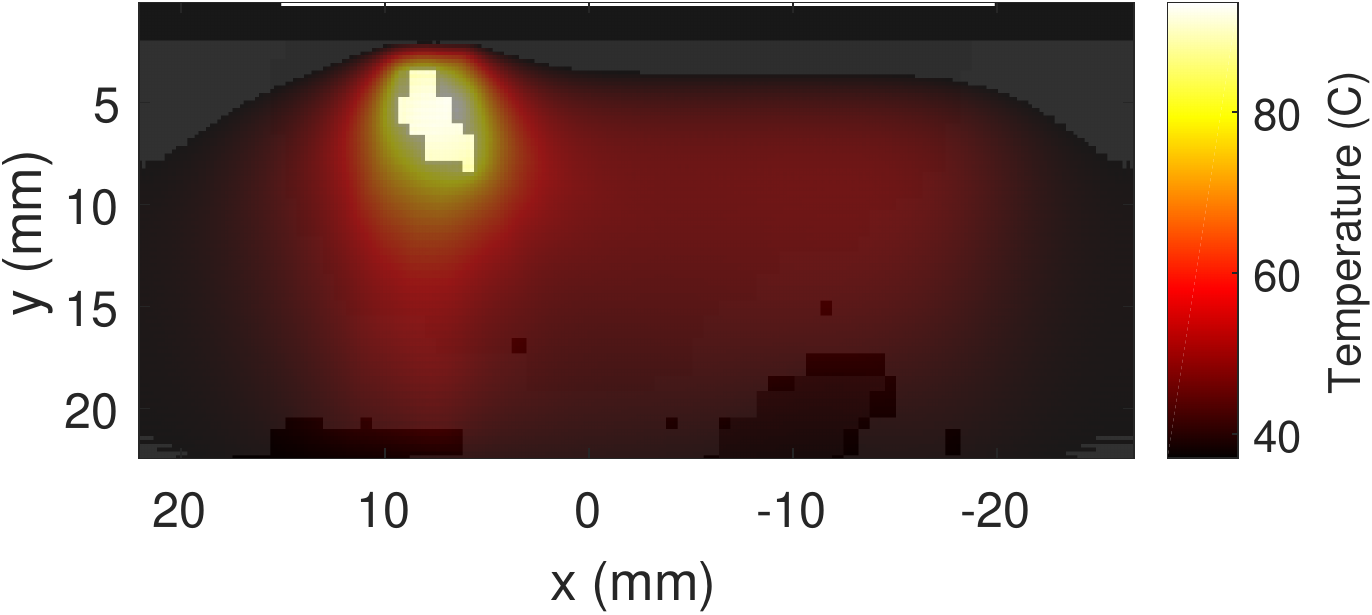}}
\caption{The execution of a 20+40-second sonication strategy in patient 2. (a) A 20-second sonication with all the elements followed by (b) a 40-second sonication using the single element in front of the calcification. (c) Temperature map at the end of the 20+40-second sonication shows the distribution of heat around the calcification.}
\label{fig:sonication_strategy}
\end{figure}

Figure~\ref{fig:sonication_strategy} shows the execution of a 20+40-second sonication strategy in patient 2. In this approach, a 20-second therapeutic sonication with all elements active was followed by a 40-second sonication using only the element(s) in front of the calcification(s) active (in this case, a single element). In Figure~\ref{fig:sonication_strategy}(b) it can be seen that the calcification effectively blocks the ultrasound field from penetrating to the rear region. The resulting temperature distribution at the end of the sonication is also shown in Figure~\ref{fig:sonication_strategy}(c). It is evident that the strongest heating is in front of the calcification from where it is conducted to the rear side and further into the prostate tissue.

\begin{figure}[b!]
\centering
\subfigure[]{\includegraphics[height=3cm]{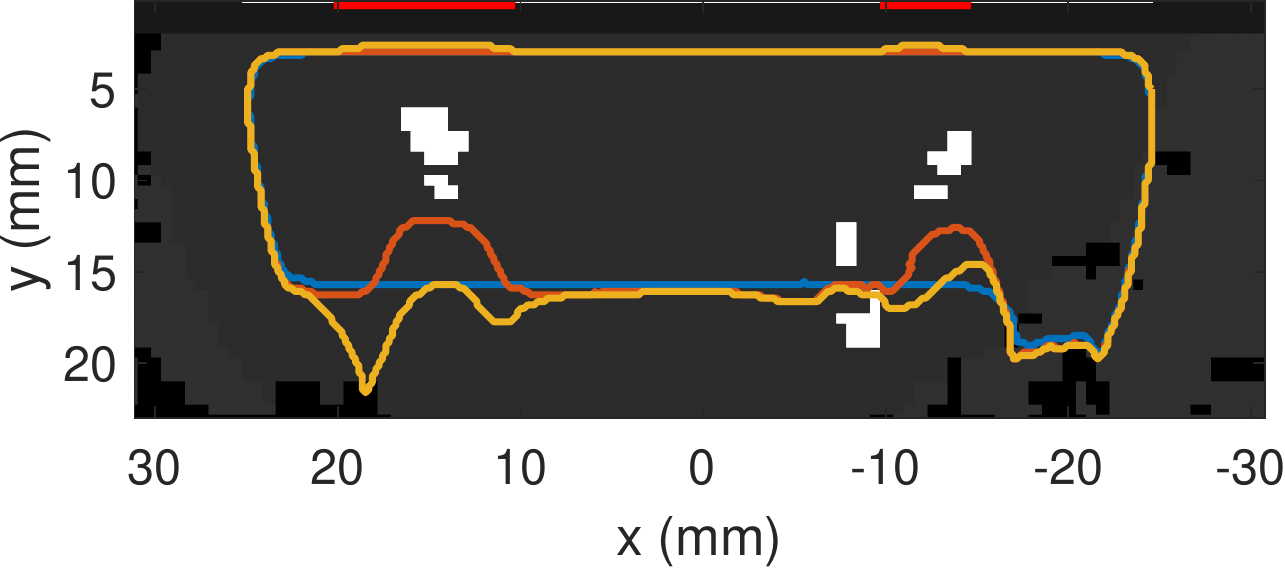}}
\subfigure[]{\includegraphics[height=3cm]{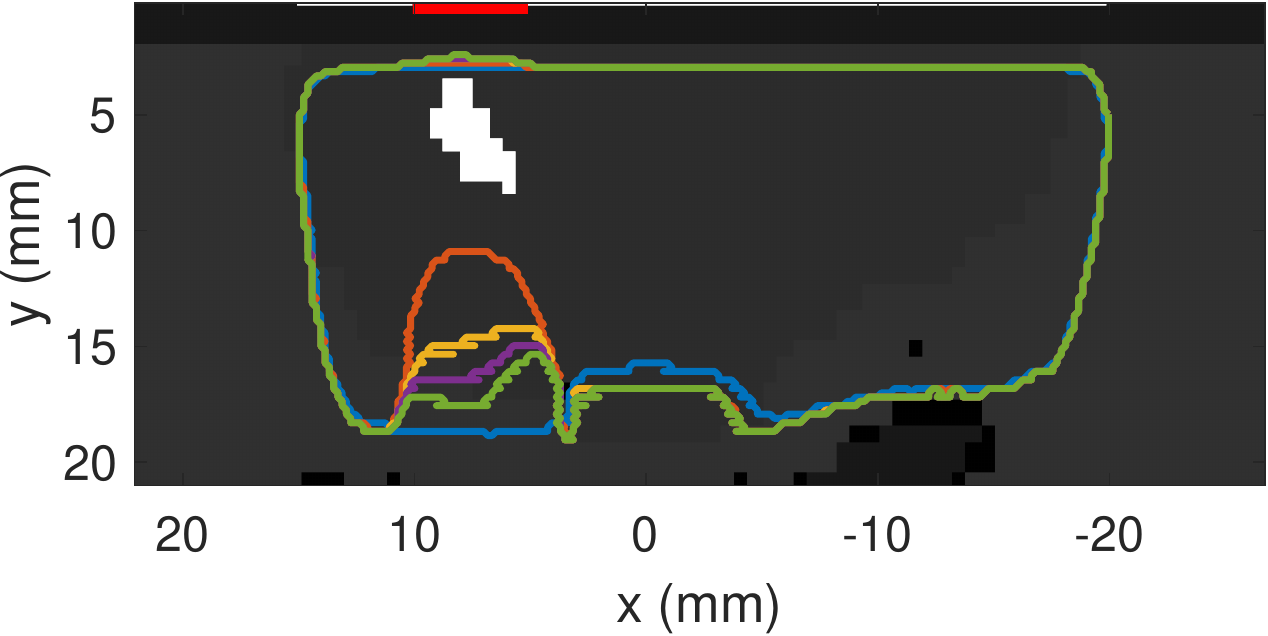}}
\subfigure[]{\includegraphics[height=3cm]{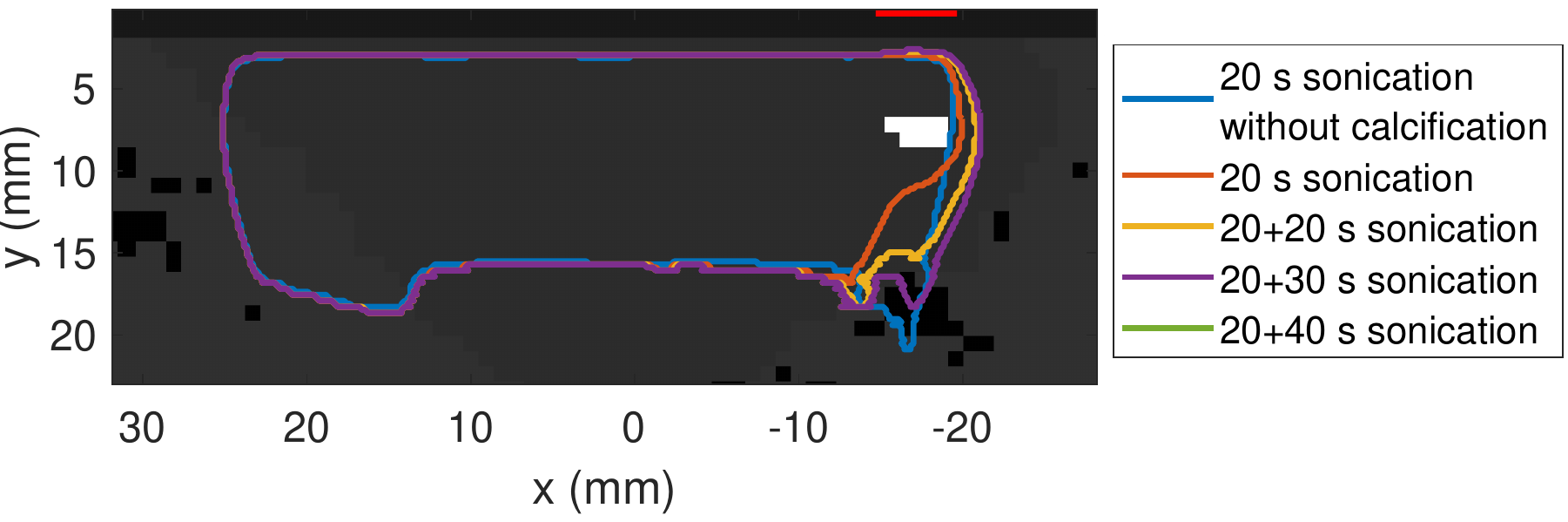}}
\caption{(a)-(c) The boundaries of 240 cumulative equivalent minutes (CEM) thermal dose areas on the axial plane cutting through the centre of the transducer in patients 1-3, respectively. Therapeutic 20-second sonications were conducted with and without the calcifications in addition to different sonication strategies. The selective elements are highlighted in red.}
\label{fig:strategy_comparison}
\end{figure}

Figures~\ref{fig:strategy_comparison}(a)-(c) show the boundaries of the thermal dose areas resulting from different sonication strategies in patients 1-3, respectively. In each case, different sonication strategies were tested in order to treat the tissue region behind the calcifications. The maximum temperature reached during the strategy sonications was 93.9~$^{\circ}$C, which is still below the boiling temperature.

Figure~\ref{fig:strategy_comparison}(a) shows the contours of the thermal dose areas on an axial plane cutting through the centre of the transducer in patient 1. In this case, there are multiple calcifications at different locations in the prostate. Conducting a 20-second therapeutic sonication leaves noticeable untreated areas behind the calcifications when compared to the same sonication without the calcifications. The 20+20-second strategy sonication shows the boundaries after a full 20-second therapeutic sonication with all the elements active followed by a 20-second sonication using only the elements 3, 8 and 9 (from right to left). The sonication strategy successfully ablates the regions behind the calcifications. However, some additional damage is also present beyond the boundary of the original sonication. This was because elements 8-9 partly overlap the edges of the calcification, and thus were able to penetrate deeper into the prostate.

Figure~\ref{fig:strategy_comparison}(b) shows the thermal dose boundaries in patient 2 with different sonication strategies. After a 20-second sonication with all elements active followed by a 40-second cooling time, the calcification has left a large untreated region behind. When the 20-second sonication was followed by a 20-second sonication with a single element (i.e., 20+20 s) the untreated region is greatly reduced (approximately by 50\%). When the sonication duration with a single element was further increased to 40 seconds, the region boundary covers almost the same extent as the therapeutic sonication without the calcification. However, caution has to be taken in the region between the calcification and the transducer, where some thermal damage has also occurred.

Figure~\ref{fig:strategy_comparison}(c) shows the thermal dose boundaries in patient 3. In this case, the calcification was located at the edge of the prostate right in front of the first transducer element. After a 20-second therapeutic sonication, the treated region at the boundary had narrowed compared to the sonication without the calcification. When the 20-second sonication was followed by a 20-second or 30-second sonications using only the first element, the treated region is increased. Some thermal damage also extended outside the prostate in the lateral (x) direction, which is not visible in the sonication without the calcification.

\section*{Discussion}

The purpose of this study was to investigate clinical questions regarding the therapeutic ultrasound treatment of the prostate. In particular: What is the maximum size of prostatic calcification that does not limit the treatment region? What is the minimum distance of the calcification from the transducer so that no thermal damage is induced in the urethra? What is the best strategy to treat the tissue region behind the calcification? In order to answer these question, realistic computational simulations were conducted using clinical patient data. The simulations were carried out with both naturally occurring and artificial calcifications whose sizes and locations were varied.

Regarding the maximum size of the calcification, the answer is somewhat expected: it depends. Even with a 1~mm diameter calcification, a small `dint' was seen at the outer boundary of the thermal dose area. Whether the size of this untreated region is clinically significant or not is another question. As the size of the calcification increased, the area of the untreated region behind the calcification grew to a certain point. However, when the size of the calcification increased sufficiently (7~mm and above), the treated region behind the calcification started growing again. This happened due to the fact that the thermal conductivity of the calcifications is relatively high, and therefore, some of the heat energy in front and inside the calcification was dissipated to the tissue regions surrounding the calcification. To conclude, even calcifications smaller than 10~mm in diameter can cause large untreated tissue regions if they are located close to the transducer. The maximum size of the calcification used in this study was 10~mm in diameter, but also larger occurrences are possible~\citep{hong2012prevalence}.

Another factor affecting the size of the untreated region behind the calcification is the distance of the calcification from the transducer. As the distance from the transducer becomes larger, the untreated region becomes smaller. At large distances, the treatment region actually expands beyond what it would be without the calcification due to thermal conductivity. Therefore, it is advisable to avoid heating calcifications which are located very close to critical structures such as the urethra, bladder neck or rectal wall.

The minimum distance at which the calcification can be located from the transducer before any thermal damage is induced in the urethra is dependent on the sonication duration and the efficiency of the heat sink, which in this case was the constant temperature ultrasound probe. In the simulations, the closest distance of the surface of the calcification from the transducer was 2.5~mm, which is only 0.5~mm away from the ultrasound probe (there exists a 2~mm layer of water between the transducer and the sonication window). At this distance, slight thermal damage (approximately 0.5~mm) extended from the calcification towards the transducer during the 20-second sonication. Only when the distance was increased to 7.5~mm, no thermal damage in the urethra was observed. Thus, heat accumulation can be expected to occur between the transducer and the calcification, which could potentially cause thermal damage to the urethra if the cooling is not sufficient. Therefore, a margin of at least 7.5~mm between the transducer and the calcification should be left in order to avoid additional thermal damage to the urethra.

With regards to the best sonication strategy to treat the tissue regions behind calcifications, two different approaches can be taken: lowering the sonication frequency or increasing the sonication duration. The former approach is based on the assumption that lower frequency ultrasound waves could potentially penetrate through calcifications with smaller energy loss due to the lower attenuation. However, it was observed that the boundary of the untreated region behind the calcification remained approximately the same when the sonication frequency was changed between 1 and 4~MHz while keeping the sonication duration constant. This occurred despite the fact that higher frequencies exhibit higher heating rate when the power level is kept constant. This suggests that a better strategy might be using longer sonication durations.

Longer sonication durations were seen to extend the boundary of the treated region behind the calcification, but at a lower pace than in the regions where there were no calcifications. Therefore, a better strategy would be to first treat the desired tissue regions and then only focus on the remaining untreated region behind the calcification. In this case, the elements directly in front of the calcifications were selected to treat the remaining untreated tissue regions behind.

The simulation results for calcifications in this study are in agreement with previous research studying ultrasound beam distortion due to fiducial markers during transrectal focused ultrasound therapy in the prostate~\citep{georgiou2017beam, bakaric2018experimental}. It was found that the beam distortion due to the markers may result in an untreated region beyond the marker due to reflections. Although the acoustic properties of gold fiducial markers are different to those of the prostatic calcifications, they both act as strong acoustic reflectors.

Temperature mapping during treatments is usually performed using proton resonance frequency shift based MRI thermometry. Unfortunately, this technique is prone to susceptibility effects that cause a signal void in temperature maps. In addition, these techniques can be very sensitive to temperature-induced tissue susceptibility changes~\citep{sprinkhuizen2010temperature}. There could be even false temperature readings in the immediate vicinity of calcifications. The severity of this problem is likely to depend on the size of calcification, but it can cause an additional challenge when validating the results of these simulations by phantom experiments or during actual treatments. On the other hand, it underlines the importance of simulation studies in order to estimate the temperature distribution around calcifications.

The simulation models used in this study were realistic taking into account all the relevant acoustic and thermal effects such as nonlinearity and perfusion. The phenomenon not taken into account that could possibly be induced by the presence of calcifications in the ultrasound field is the acoustic cavitation~\citep{coussios2007role}. The strong acoustic reflections caused by the calcifications create a large negative pressure in front of the calcification, which could potentially trigger cavitation in the same location. In the simulations, the maximum mechanical index (MI) in front of the calcifications was 1.0 which is well below the Food and Drug Administration (FDA) recommended limit of 1.9 for diagnostic ultrasound devices~\citep{szabo2004diagnostic}. Therefore, it is unlikely that cavitation would cause any tissue destruction in the region in front of the calcification. 

\section*{Conclusions}

The aim of this study was to examine how prostatic calcifications of different sizes and locations affect the efficacy of transurethral ultrasound therapy in the prostate. It was found that for fixed sonication parameters, even small calcifications leave untreated regions behind them due to strong acoustic reflections and high attenuation. Furthermore, increased temperature elevation between the calcification and the transducer was seen.

In addition, different sonication strategies were tested in order to deliver therapy to the untreated regions behind the calcifications. The best strategy was to conduct prolonged sonications using the elements directly in front of the calcifications. This way the heat energy was conducted through the calcification to the surrounding tissue regions. However, attention should be paid to avoid calcifications located close to sensitive tissues such as the urethra, bladder neck or rectal wall.


%

%

\section*{Acknowledgements}

V.~S. acknowledges the support of the State Research Funding (ERVA), Hospital District of Southwest Finland, number K3007 and CSC - IT Center for Science, Finland, for computational resources. J.~J. was supported by The Ministry of Education, Youth and Sports of the Czech Republic from the National Programme of Sustainability (NPU II); project IT4Innovations excellence in science - LQ1602. B.~T. acknowledges the support of EPSRC grant numbers EP/L020262/1 and EP/M011119/1.





\pagebreak

\bibliographystyle{UMB-elsarticle-harvbib}
\bibliography{arXiv_MedPhys}

\end{document}